\documentclass[sigconf]{acmart}

\usepackage{graphicx}
\usepackage{array}
\usepackage{booktabs}
\usepackage{longtable}
\usepackage{xcolor}
\usepackage{multirow}

\newcolumntype{C}[1]{>{\centering\arraybackslash}m{#1}}

\newcommand{\revise}[1]{\textcolor{black}{#1}}

\AtBeginDocument{%
  }

\copyrightyear{2025}
\acmYear{2025}
\setcopyright{cc}
\setcctype{by}
\acmConference[CHI '25]{CHI Conference on Human Factors in Computing Systems}{April 26-May 1, 2025}{Yokohama, Japan}
\acmBooktitle{CHI Conference on Human Factors in Computing Systems (CHI '25), April 26-May 1, 2025, Yokohama, Japan}\acmDOI{10.1145/3706598.3713724}
\acmISBN{979-8-4007-1394-1/25/04}

\begin{document}

\title{What Social Media Use Do People Regret? An Analysis of 34K Smartphone Screenshots with Multimodal LLM}

\author{Longjie Guo}
 \affiliation{%
   \department{Information School}
   \institution{University of Washington}
   \city{Seattle}
   \state{Washington}
   \country{USA}}
 \email{longjie@uw.edu}

  \author{Yue Fu}
  \affiliation{%
   \department{Information School}
   \institution{University of Washington}
   \city{Seattle}
   \state{Washington}
   \country{USA}}
 \email{chrisfu@uw.edu}

  \author{Xiran Lin}
 \affiliation{%
   \department{Global Innovation Exchange}
   \institution{University of Washington}
   \city{Bellevue}
   \state{Washington}
   \country{USA}}
 \email{xiranlin@uw.edu}

   \author{Xuhai "Orson" Xu}
 \affiliation{%
   \department{Department of Biomedical Informatics}
   \institution{Columbia University}
   \city{New York City}
   \state{New York}
   \country{USA}}
 \email{xx2489@columbia.edu}

    \author{Yung-Ju Chang}
 \affiliation{%
   \department{Department of Computer Science}
   \institution{National Yang Ming Chiao Tung University}
   \city{Hsinchu}
   \country{Taiwan}}
 \email{armuro@nycu.edu.tw}

  \author{Alexis Hiniker}
  \affiliation{%
   \department{Information School}
   \institution{University of Washington}
   \city{Seattle}
   \state{Washington}
   \country{USA}}
 \email{alexisr@uw.edu}

\renewcommand{\shortauthors}{Guo et al.}

\begin{abstract}
Smartphone users often regret aspects of their phone use, \revise{especially social media use. However,} pinpointing specific ways in which the design of an interface contributes to regrettable use can be challenging due to the complexity of \revise{social media} app features and user intentions. We conducted a one-week study with 17 Android users, using a novel method where we passively collected screenshots every five seconds, which we analyzed via a multimodal large language model to understand participants' usage activity at a fine-grained level. Triangulating this data with data from experience sampling, surveys, and interviews, we found that regret varies based on user intention, with non-intentional and social media use being especially regrettable. Regret also varies by social media activity; participants were most likely to regret viewing algorithmically recommended content and comments. Additionally, participants frequently deviated to browsing social media when their intention was direct communication, which slightly increased their regret. Our findings provide guidance to designers and policy-makers seeking to improve users’ experience and autonomy.
\end{abstract}




\maketitle

\section{Introduction}
Despite the popularity of \revise{social media platforms}, smartphone users consistently say that they engage with social media in ways they later regret. They report scrolling through content that is not worth their time and checking for updates compulsively~\cite{tran2019, cho2021}. This regrettable usage is no accident; in the attention economy, online platforms are \revise{often} intentionally designed to keep users on task for as long as possible and to manufacture routine phone-checking habits that keep them coming back as often as possible~\cite{eyal2014hooked}. An increasing body of prior work argues that the concept of designing for ``engagement'' has received too little scrutiny and deserves greater attention from regulators~\cite{richards2024against}. \revise{This is particularly true of social media platforms, which are frequently built around attention-economy business models and provide the usage experiences that users are most likely to say they regret~\cite{tran2019, cho2021}.}

Yet, understanding the precise relationship between users' engagement with \revise{social media apps on the phone} and their subsequent regret is incredibly challenging. First, users can engage in a variety of activities throughout their interaction in a given app, as many mobile apps today consist of a complex constellation of features. A single social media app might include, for example: a feed of algorithmically organized content, direct messaging features, discussion threads, and more, making it difficult to associate isolated design decisions with users' usage decisions or subjective experiences. Indeed, prior research has examined the relationship between feature-level social media app usage and subsequent feelings of regret~\cite{cho2021} and showed that regret of social media usage can vary by feature. 

Second, people bring many different intentions to their app use, and prior work suggests a user's motivation for engaging with an app may influence whether they later regret doing so~\cite{lukoff2018, hiniker2016}. %
For instance, earlier studies have shown that users often turn to mobile apps to kill time during idle moments~\cite{chen2023, pielot_when_2015, zheng2024fragmented, yang2020bored}, typically without a clear purpose. It is plausible that users' sense of regret after such mindless use differs from their feelings after engaging with the same app for specific, goal-directed tasks, such as communicating with a friend or seeking information. Finally, prior work suggests that whether a user's actual activity aligns with their intention may also influence regret~\cite{cho2021, zeelenberg2007, pieters2005}. 
Thus, in this study, we sought to answer the following research questions: 

\begin{itemize}
    \item RQ1: How does regret vary based on users’ different intentions for app use?
    \item RQ2: How does regret vary based on the specific activities users engage in on social media apps?
    \item RQ3: How does regret vary based on whether users' actual activities \revise{on social media apps} align with their intention?
\end{itemize}

To answer these research questions, we conducted a \revise{one-week,} mixed-methods study with 17 Android users. \revise{We examined participants' phone use to enable us to understand their use of social media apps---and any regret it produces---in the larger context of all phone use. By documenting all forms of phone use and measuring associated regret (if any) across a variety of usage experiences, we were able to compare social media regret with that of other use cases.} Specifically, we used the experience sampling method to capture user intentions in real-time, combined with automatic and continuous screenshot capturing every five seconds. The collected screenshots were then analyzed by a multimodal large language model~\cite{openai2024gpt4o}, which is capable of interpreting visual content and generating descriptive information. We used this novel approach to systematically and efficiently analyze fine-grained phone-use behaviors in social media apps, overcoming the scalability challenges of manually coding screenshots exhaustively and the lack of flexibility of alternative approaches, such as directly retrieving UI components of an app~\cite{cho2021, orzikulova2023}. \revise{At the end of each day, participants reviewed screenshots of their own usage and labeled the sessions that they regretted. This gave us the opportunity to examine short-term, action-based regret (which prior work differentiates from actions one \textit{failed} to take~\cite{gilovich1995}).} We then complemented these \revise{daily annotations} with retrospective interviews, where users reflected on their phone use over the course of the study.

\revise{Our results first showed} that participants' regret varied based on intended use, with non-intentional use being most regretful, and that participants regretted social media use more than other types of phone use. \revise{Then, through our LLM-powered analysis of 34,313 smartphone screenshots, we found that on social media apps}, regret varied based on social media activity, and that participants found viewing algorithmically recommended content and comments most regretful. \revise{Comparing participants' initial intentions \revise{for using social media apps} with the activities they ultimately engaged in reveals that participants frequently intended to communicate but ended up engaging in other social media activities instead (over 60\% of the time). Additionally, participants expressed slightly more regret for sessions where they deviated from their original intentions compared to sessions where they did not.} Our regression model showed that duration of app use, user intention, and proportions of recommendation- and subscription-based content and comments are significant indicators of regret. By investigating both intention and fine-grained social media activity and how they are related to regret, we provide guidance for designers and policy-makers to design experiences that respect users' autonomy and intentions.

\section{Related Work}
\revise{Smartphone users are presented with an infinite number of usage choices, including which apps to use, how to engage with them, and what media content to consume. Further, many platforms (especially social media platforms) rely on attention-economy business models, which has led to designs that pressure users into spending as much time with the product as possible~\cite{roffarello2023}. This combination of information overload and pressure from platforms creates a context where users, at times, feel regret about their technology use and realize that their time would have been better spent elsewhere~\cite{tran2019}. Because regret is predominantly experienced as a negative and aversive emotion~\cite{zeelenberg2007}, understanding and regulating it is highly consequential to digital well-being. Here, we describe the prior strands of research that we draw on in examining some of these regrets. This includes: prior work studying the technology use that people find problematic, interventions to address this problematic usage, technical methods for studying problematic usage behaviors, and theoretical conceptualizations of regret.}

\subsection{Understanding Problematic Smartphone Use}
Although the widespread adoption of smartphones has brought numerous benefits, their increasing ubiquity and utility have also heightened users' attention to them. Recent research shows that many smartphone users now check their phones every few minutes, with a significant portion of these interactions being self-initiated~\cite{dingler2015ll,pielot2014situ, heitmayer2021smartphones}. A large body of research has investigated user's experience with problematic phone use~\cite{lukoff2018, tran2019, cho2021, oulasvirta2012habits, chan2015, elhai2017, fox2015}. For example, Chan used a survey to examine the relationship between phone use and subjective well-being and found that communicative uses of mobile phone are positively related to subjective well-being while non-communicative uses are negatively related to subjective well-being~\cite{chan2015}. Tran et al. found that compulsive phone use might be triggered by unoccupied moment, tedious task, social awkwardness, and anticipation, and that users express frustration with such compulsive checking habits, unless the checking behavior resulted in experiences that they find meaningful and that transcend phone use, such as relationship-building~\cite{tran2019}.

One insight coming out of this body of work is that understanding phone use at the app level or app category level may be insufficient and too coarse-grained, since many mobile apps today, especially social media apps (which are most likely regrettable~\cite{cho2021, tran2019}), offer a variety of features, where user behavior might vary depending on the feature~\cite{cho2021} or depending on how they intend to use the app~\cite{lukoff2018}. Drawing on the Uses and Gratifications theory, Lukoff et al. investigated what smartphone use is meaningful and meaningless to people by asking for their motivation and type of phone use, and found that people feel a lower sense of meaningfulness when their phone use is motivated by habitual use to pass the time and when people use their phone for entertainment and passively browsing social media~\cite{lukoff2018}.  By using a feature-level analysis approach and incorporating the construct of regret, Cho et al. found that users felt more regretful about social media features that comprise passive forms of usage (such as viewing social media feed) than active forms of usage (such as searching and messaging), and that they regretted habitual checking on their feed, sidetracking from original intention to recommendation-based features, and falling into prolonged use when viewing recommendation-based content~\cite{cho2021}. In this study, we incorporate these perspectives, and combine the experience sampling method and fine-grained understanding of user's moment-to-moment behavior via passively collected screenshots to investigate how regret varies based on what motivates people to use their phone (their intention) and what they subsequently do once they are on their phone, offering more nuanced perspectives into regrettable phone-use behavior.

While much digital well-being and problematic smartphone use research seeks to understand smartphone users and design individual-oriented solutions to reduce problematic phone use, there are also exogenous factors underlying technology overuse, as technology is often intentionally designed to optimize engagement and nudge users towards problematic use in the attention economy~\cite{eyal2014hooked, fasoli2021overuse}. An increasing body of prior work has argued that the concept of designing for ``engagement'' in technology platforms deserves more scrutiny from regulators~\cite{richards2024against, europeanparliament2023digitaladdiction, atlantic2016bingebreaker}. Richards \& Hartzog outlined several harms associated with the engagement model of digital platforms, including privacy violations, the erosion of attention (coined as ``attention theft''), and detrimental impacts on mental health, relationships, and democratic processes, and argue that wrongful engagement strategies should be regulated~\cite{richards2024against}. Our study provides insights into how the design of digital platforms aiming at maximizing engagement can create regrettable experiences for their users and sheds light on how policy-makers can help improve user's experience and autonomy.

\subsection{Problematic Smartphone Use Interventions}
 Due to the negative effects of smartphone use, many tools and intervention mechanisms aiming at reducing smartphone use have been implemented. Both Android and iOS have a default system-level tool to track app usage and set limits~\cite{apple2024screentime, google2024wellbeing}. Researchers have also demonstrated the effectiveness of various intervention techniques to reduce smartphone use~\cite{hiniker2016mytime, lu2024interactout, xu2022, wu2024, orzikulova2023, orzikulova2024time2stop, kim2019}. For example, MyTime uses three mechanisms: timer, timeout, and aspiration to help decrease overuse~\cite{hiniker2016mytime}. InteractOut leverages input manipulation techniques to inhibit natural user gestures on smartphones to reduce overuse (e.g., by delaying tap and swipe)~\cite{lu2024interactout}. TypeOut leverages self-affirmation (letting users type statements like ``I value self-control'' when entering an app) to reduce smartphone overuse~\cite{xu2022}. MindShift utilizes large language model to adaptively tailor intervention message based on user's mental state~\cite{wu2024}. However, most of these interventions are applied at the granularity of the app or device level, where once an app is opened or a usage limit is surpassed, the proposed intervention mechanism would discourage users from continuing to use the phone. Recent work has also implemented more fine-grained feature-level interventions, where users can set limits on specific features inside social media apps (e.g., consuming content on social media feed), which has been shown to decrease passive usage related to content consumption more than app-level intervention~\cite{orzikulova2023}. In addition to mechanisms that use time or app launch as triggers, researchers also proposed intervention mechanisms that use machine learning models to identify best timing for intervention based on smartphone sensor and log data to deliver just-in-time intervention~\cite{orzikulova2024time2stop}.
 
 While we do not propose a new intervention design in this paper, our empirical findings can inform the design of intervention techniques that consider user intention and fine-grained activity to reduce problematic phone use. Our screenshot analysis approach can also inform intervention systems that aim to leverage multimodal data to deliver fine-grained and just-in-time intervention.

 \subsection{Capturing Smartphone Screenshots}
Past work on understanding phone-use behaviors has mostly utilized methods such as interviews (e.g., \cite{tran2019}), phone log data (e.g., \cite{hiniker2016, falaki2010, anind}), and experience sampling data (e.g., \cite{lukoff2018}). While these approaches have helped yield insights about phone-use behavior, they can be limited since they do not reveal user's moment-to-moment behavior precisely. Some researchers have also utilized the Android Accessibility API to detect specific social media features, but this approach requires access to an app's UI structure and manual coding of all features, making it hard to scale and not robust against app updates~\cite{cho2021}. Researchers have also turned to smartphone screenshots, which provide richer information about user's behavior. Reeves et al. developed the Screenomics framework, where a system can capture and analyze personal experiences through passively collected screenshots every five seconds~\cite{reeves2021screenomics}. \revise{In addition to understanding user behavior, screenshots can also be used to predict user behavior.} Yang et al. demonstrated that screenshots collected every five seconds can be used to predict task switching~\cite{yang2019}. Chen et al. leveraged both screenshot data and sensor data to create a fusion model to predict time-killing behavior on smartphone~\cite{chen2023}. \revise{The present study primarily focuses on understanding phone use behavior through screenshots (and how certain behaviors correlate with regret), not predicting certain behavior from screenshots.}

While screenshot data captured in periodic intervals reveal rich information about user's behavior, one challenge of using such data is that they are hard to analyze. Recent advances in multimodal large language models (MLLMs) such as GPT-4o \revise{(and its predecessor, GPT-4V) have shown exceptional multimodal understanding capabilities across various domains and tasks~\cite{openai2024gpt4o, yang2023dawnlmm}. These models extend the capabilities of large language models (LLMs) by integrating multiple modalities (most typically, vision and language), and prior research has demonstrated that these systems excel across a spectrum of tasks, from simpler ones such as open-ended image description and object localization to more complex challenges such as understanding multi-image sequences and navigating graphical user interfaces (GUIs)~\cite{yang2023dawnlmm}. The GUI understanding capability of MLLMs is of particular interest to understanding phone-use behavior, as researchers have demonstrated that, based solely on screenshots, MLLMs can understand the visual content of mobile UI screens (e.g., summarizing content or activity from screenshots) across different apps, and even operate these apps by predicting future actions~\cite{wang2024mobileagentautonomousmultimodalmobile, zhang2023appagentmultimodalagentssmartphone, yang2023dawnlmm}. Other studies demonstrated that MLLMs can automatically generate code based on screenshots~\cite{wan2024automaticallygeneratinguicode, yang2023dawnlmm}. Such capabilities make MLLMs an ideal candidate to automatically analyze phone-use data recorded in screenshots, which can overcome the challenge of manual labeling and provide flexibility through prompting.} In this study, we explore this possibility and introduce a novel method to automatically categorize social media use into fine-grained activities using GPT-4o.

 \revise{\subsection{Theorizing Regret}} 
 \revise{A large body of literature in economics and psychology focuses on operationalizing and characterizing regret~\cite{bell1982, miller1990, zeelenberg1999anticipated, zeelenberg2007, gilovich1995, pieters2005}. First, there is a general consensus that, unlike many other emotions, regret is a \textit{cognitive} emotion, where thinking and judgment are central~\cite{gilovich1995,zeelenberg2007}. In economics, regret is defined as ``\textit{the consequence of decision-making under uncertainty},'' a reaction to the simple difference between experienced reality and rejected alternatives~\cite{bell1982}. Research on counterfactual thinking highlights that the outcome of counterfactual realities can be imagined (rather than already known) and that the path by which a decision is made can also influence regret~\cite{miller1990}.} 
 
 \revise{In Gilovich and Medvec's seminal work on regret, the authors divide regret into regret of \textit{action} (regretting something one has done) and \textit{inaction} (regretting that one has not done something), and show that there is a temporal pattern to regret---namely that actions produce more regret in the short term while inactions produce more regret in the long term~\cite{gilovich1995}. In other words, people experience more intense, immediate pain from a regrettable action, but over the longer arc of their life, they experience more enduring distress as a result of the ways in which they have failed to act.} 
 
\revise{Zeelenberg and Peters define regret as ``\textit{an aversive, cognitive emotion that people are motivated to regulate}'' and ``\textit{a comparison-based emotion of self-blame, experienced when people realize or imagine that their present situation would have been better had they decided differently in the past}''~\cite{zeelenberg2007}. They also explain that regret can be divided into \textit{process regret} (regret that stems from a poor decision-making process) and \textit{outcome regret} (regret that stems from dissatisfaction with an outcome), and that whether a decision is \textit{justifiable} can play an important role in determining regret, independent of the decision outcome~\cite{zeelenberg2007}. For example, intention-behavior inconsistency (not behaving in ways originally intended) can amplify regret, since behaviors that deviate from the original intention are often less justifiable, which exemplify poor decision-making processes~\cite{zeelenberg2007, pieters2005}. Importantly, prior work has argued that regret is not a unitary emotion and that the conceptual boundaries of regret are not always clear~\cite{gilovich1995}. For example, the source of regret can vary from contexts ranging from moral transgressions to failures of self-actualization~\cite{gilovich1995}.}

 \revise{In this paper, we focus on action-based regret (and specifically, regret caused by using social media). Since this form of regret often manifests in the short term (as opposed to inaction-based regret, which is more likely to affect people over the course of their entire life span), we assessed people's regret at the end of each day. Given that regret is not a unitary emotion and can have rich meanings, we use a rather inclusive and minimally restrictive working definition in our study, similar to prior work~\cite{gilovich1995}. We asked participants whether they agreed that they felt regret about a phone use session. Our data collection method did not define regret for participants, leaving room for them to express process regret, outcome regret, or a combination of both.}

\section{Method}
To investigate the \revise{social media} use people regret, we conducted a one-week study, which consisted of an initial interview, a week of data collection in-the-wild (which included experience sampling method (ESM), passive screenshot collection, and daily questionnaires), and two follow-up interviews. Prior to launching the study, we piloted the data-collection tool and interview materials with three participants to refine the tool and materials.

\subsection{Participants}
\subsubsection{Recruitment}\label{recruit} We recruited 17 adults from social media channels (including X, Slack, Facebook groups, LinkedIn, Discord, and WeChat), university mailing lists, and on-campus fliers. All interested participants first responded to a screening survey after seeing a recruitment ad titled ``Seeking Android Users for a Paid Research Study.'' The screening survey consisted of multiple-choice and open-ended questions about their phone and phone use, the extent to which they wanted to change their phone-use habits, their motivation for joining the study, availability, potential privacy concerns regarding screenshot collection, and demographic information. We reached out to survey respondents who provided high-quality responses to the open-ended questions, reported using an Android phone as their main device with the Android version no earlier than Android 10 (which our custom-built data-collection app requires), and had access to Wi-Fi or unlimited cellular data (to make sure that they could upload data). We met with 20 potential participants for an initial interview. One was deemed ineligible to participate because they were not physically located in the U.S. The other two voluntarily decided not to participate, with one concerned about privacy, and another one unable to commit due to their schedule. The remaining 17 participants all finished the one-week in-the-wild data collection and the two follow-up interviews. All of the participants completed the study between July and August 2024.

\subsubsection{Demographics} Among the 17 participants who finished the study, 4 identify as man, 12 identify as woman, and 1 identify as non-binary person. 3 of them reported they were between 18 and 24 years old, 7 between 25 and 34 years old, 5 between 35 and 44 years old, and 2 between 45 to 54 years old. In terms of ethnicity, 8 identify as White, 5 as Asian, 2 as Black or African American, 1 as Hispanic, Latino, or Spanish, and 1 as ``Other.'' In terms of level of education, 2 of them reported having some college education, 5 having a 4-year degree, 5 having a professional degree, and 5 having a doctoral degree.

\subsubsection{Compensation} The 17 participants who completed the study all received a US\$200 Amazon gift card as a compensation. Two of the three participants who did not continue after the initial interview but finished the initial interview each received a \$10 Amazon gift card. The three pilot participants received a \$20, \$20, and \$60 Amazon gift card respectively, based on the number of days they participated in the pilot study (\$20 for each day).

\subsection{Procedure}

\subsubsection{Initial Interview} Prior to data collection, we met with each potential participant to conduct an initial interview. In the interview, we first explained the procedure of the study (including how frequently screenshots will be taken and our privacy-protection mechanisms), and asked participants general questions about their phone use. We then showed them the ESM survey questions we would be using for the study and assessed whether they were able to follow the instructions and understand the differences between the categories in the survey (details in Section~\ref{data-collection}). We invited participants who seemed interested and committed to join the study. For participants who consented to continue with the study, we helped them install our custom-built data-collection app (including granting the necessary permissions such as screen recording) and provided a short tutorial on how to use the app. Three of the initial interviews were conducted in-person, and the rest were done via Zoom.

\subsubsection{In-the-Wild Data Collection} \label{data-collection} After the initial interview, each participant then went through the data collection process for \revise{seven days, which, as prior work suggested, is ``likely to yield a fairly representative sample of the various activities individuals engage in''~\cite{esm-acm, hektner-esm-book}, and is the duration used by one of the initial experience sampling studies~\cite{csikszentmihalyi1977ecology}.} We used ESM in combination with passive screenshot collection and daily questionnaires. During the week, our custom-built data-collection app ran in the background on their phone, and selectively asked about their intended use of different apps and took screenshots every five seconds in selected sessions. \revise{Although our study primarily focuses on regret in the context of social media, we still sampled other general phone use, which we used to compare with social media use.}

\revise{To minimize prompting participants and taking screenshots too frequently, we decided to sample a subset of users' phone use sessions rather than capturing all sessions. To diversify the sampled phone sessions, we adopted an algorithm that increases the likelihood of sampling a screen session (defined as the period between the screen is turned on and off) as the elapsed time since the last sampled screen session grows. The sampling intervals used were 10 and 120 minutes. Specifically, the likelihood of selecting a session within 10 minutes of the previously sampled session was set to 0, and this likelihood increased over time, reaching 100\% after 120 minutes. In other words, the longer the phone went on without sampling a session, the more likely it was to sample the next detected session, and vice versa.} For example, when the time difference is 65 minutes, there is a 50\% chance of selecting the current screen session. \revise{The 10-minute interval was chosen based on the assumption that phone sessions occurring within this time frame are likely to share a similar context, which would conflict with our goal of sampling sessions across diverse contexts. The 120-minute interval was set so that we were guaranteed to get some sampled sessions throughout each day.}

\begin{figure*}[h!]
  \centering
  {\includegraphics[width=0.23\textwidth]{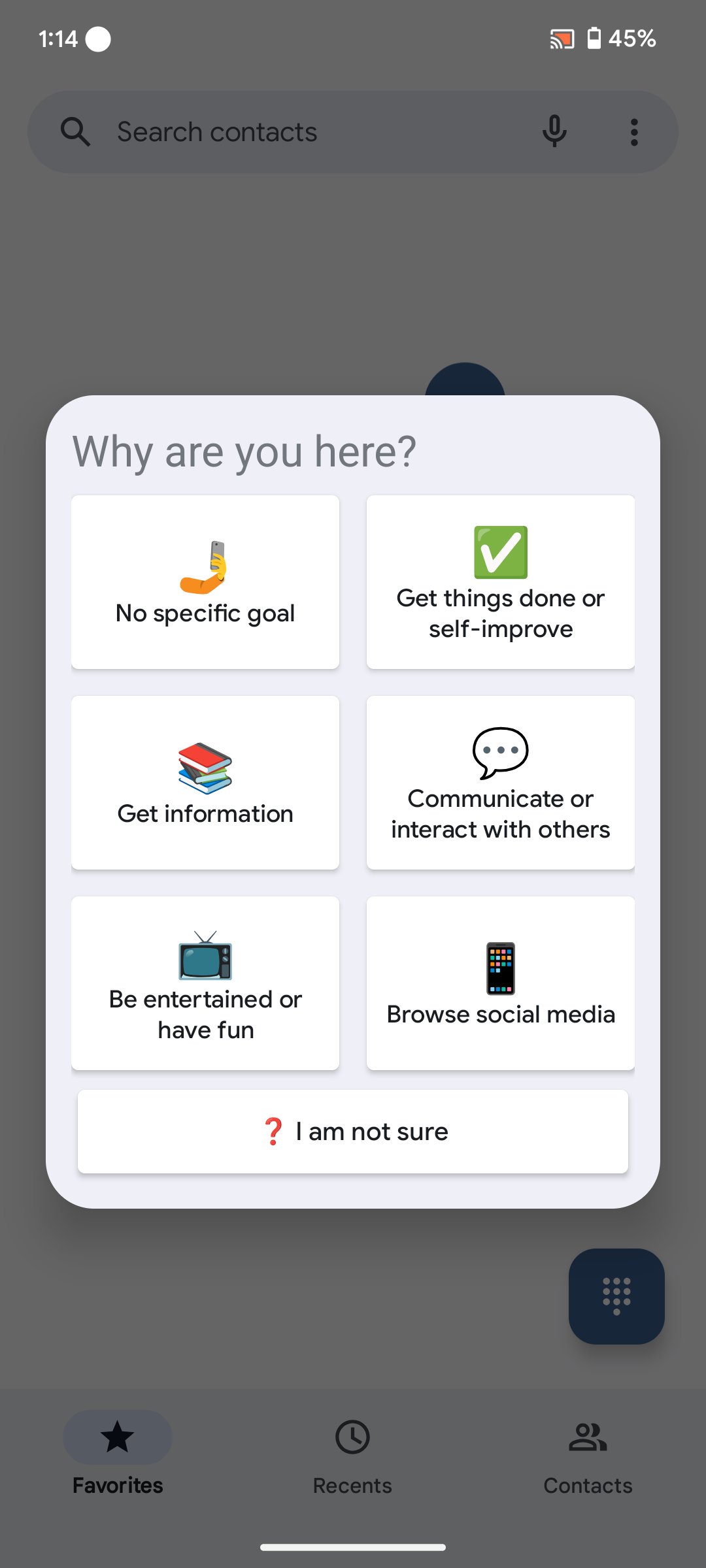}}\hfill
  {\includegraphics[width=0.23\textwidth]{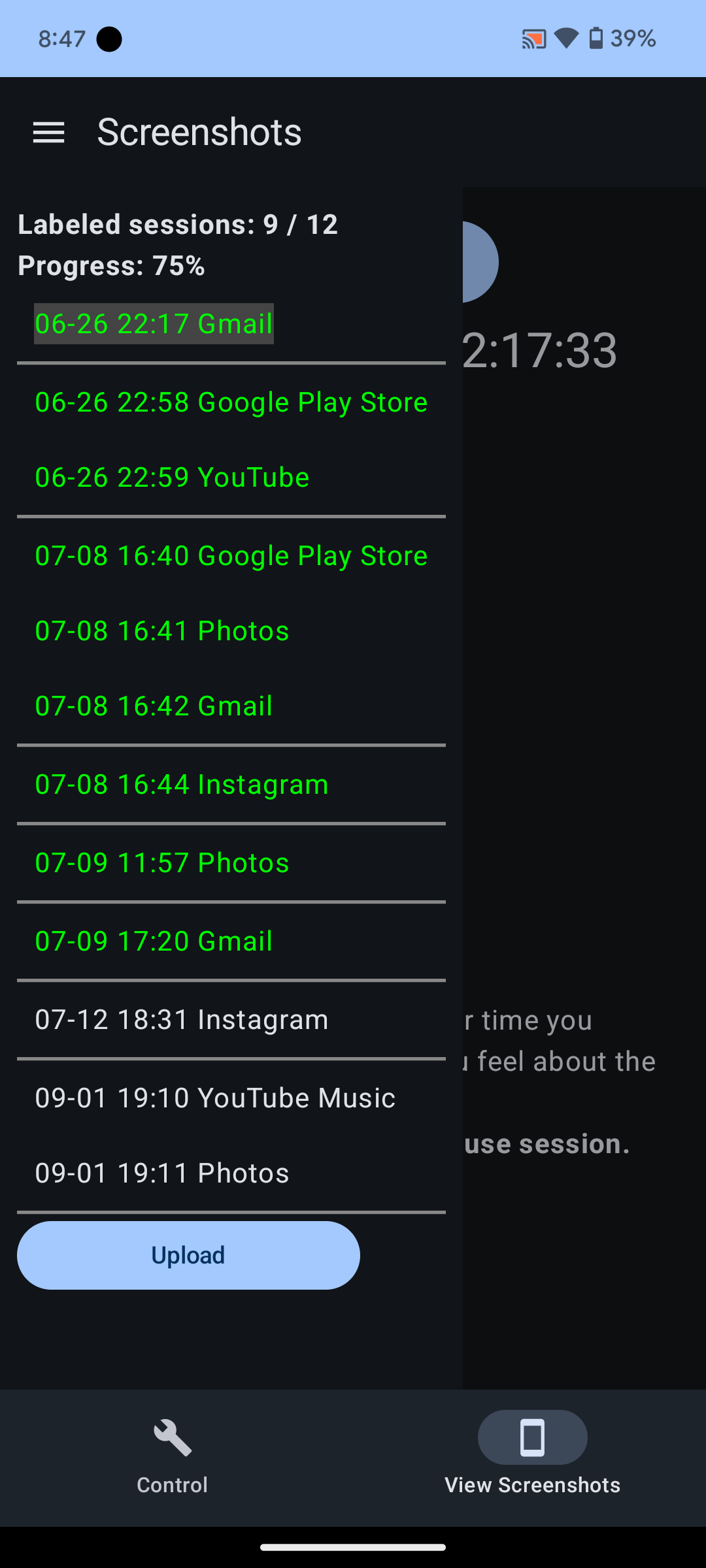}}\hfill
  {\includegraphics[width=0.23\textwidth]{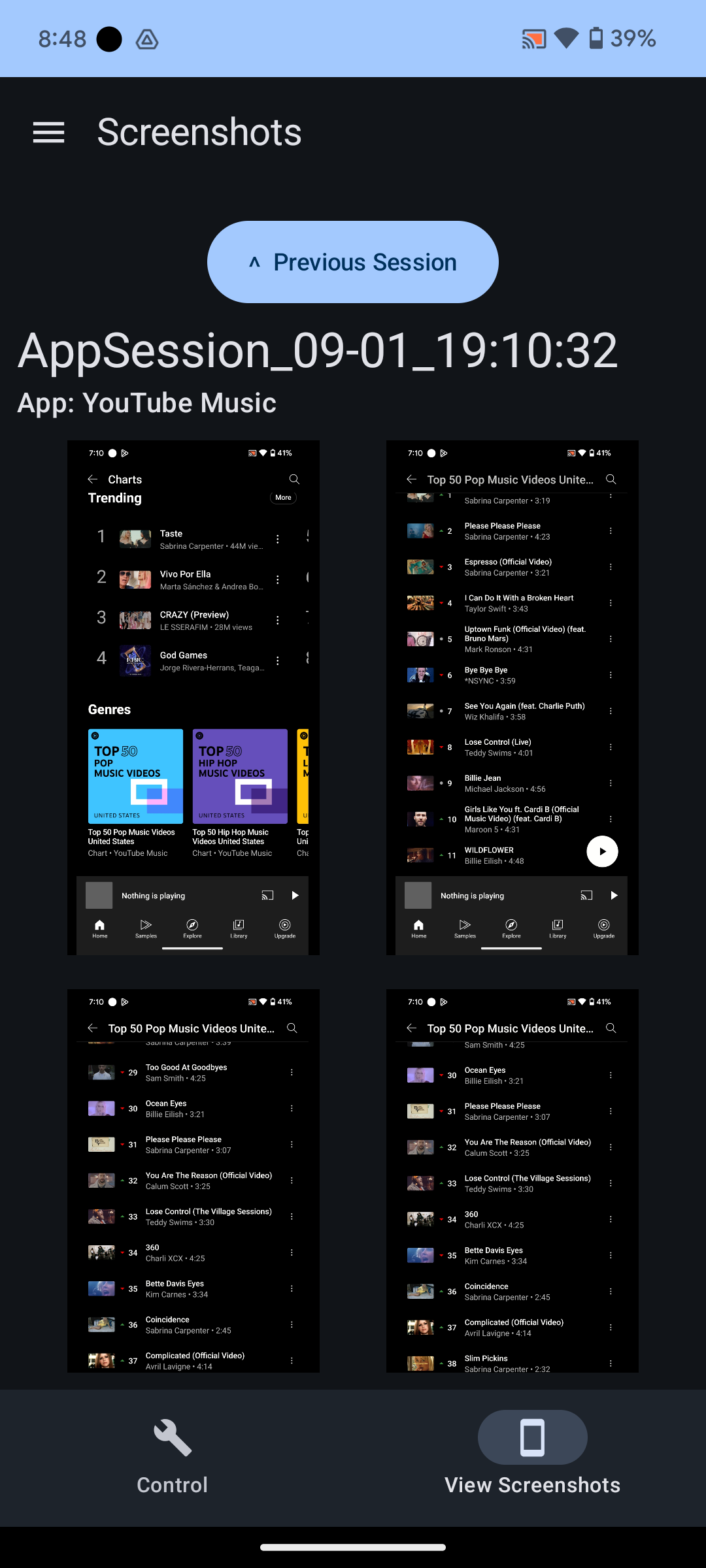}}\hfill
  {\includegraphics[width=0.23\textwidth]{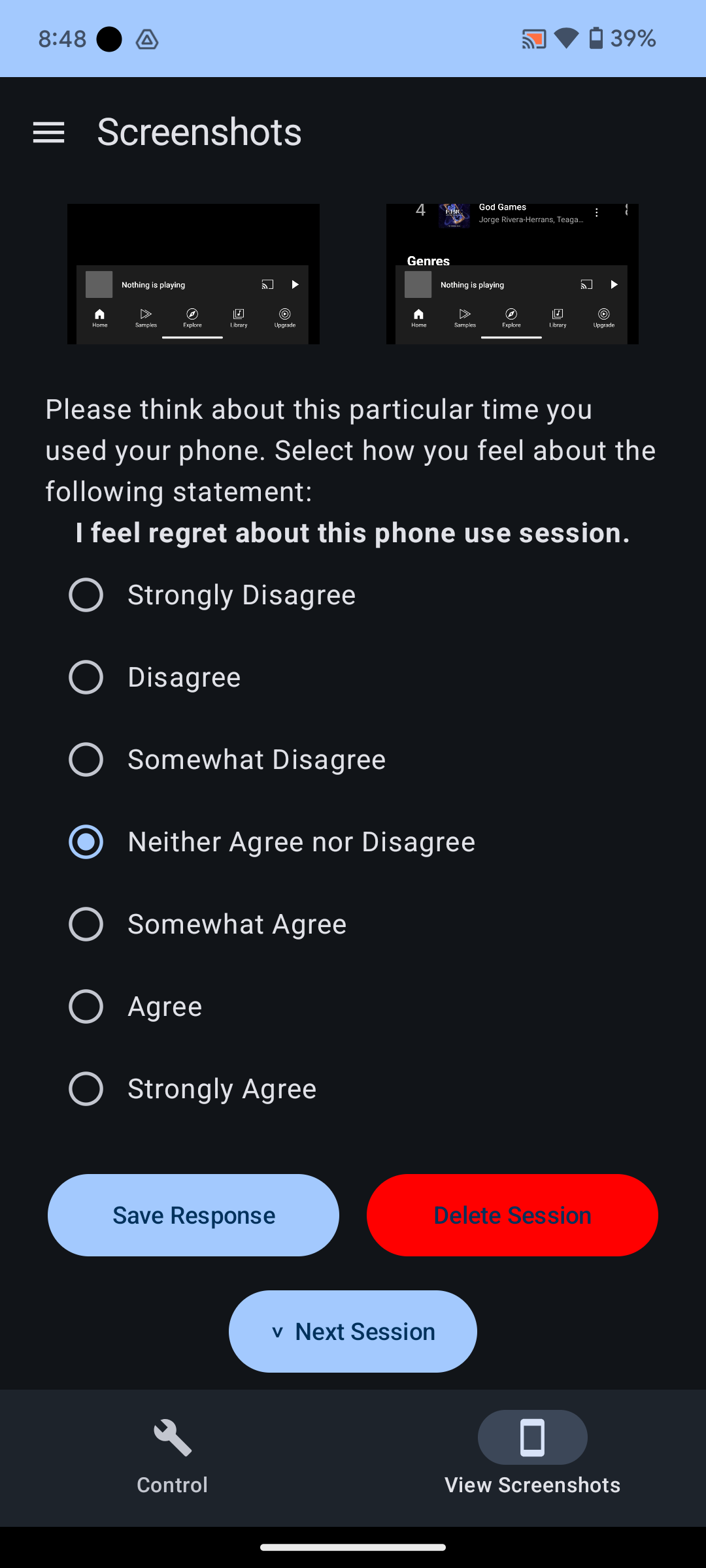}}
  \caption{Screenshots of the data-collection app. The first screenshot shows the ESM prompt asking the user about their intended use when entering each app. The second screenshot shows the interface for selecting app sessions to share with the research team and uploading them (green sessions indicate sessions where the participant finished answering the regret survey question). The third screenshot shows the app session page, which presents screenshots for each captured session, along with the starting time and app name. The last screenshot shows the regret survey question, presented at the end of each app session page.}
  \label{fig:app}
  \Description{(Left) A pop-up menu prompts users to select their reason for phone use. (Middle-left) A progress screen displays labeled phone use sessions, with timestamps and associated apps. (Middle-right) A session review screen presents a visual summary of screenshots taken during a specific app session. (Right) A survey interface asks users to reflect on their session by rating their agreement with a regret-related statement. Buttons allow users to save responses, delete sessions, or navigate to the next session.}
\end{figure*}

If a screen session was selected, \textit{every} time the participant opened or switched to a new app (we define the period an app stays on the foreground as an \textit{app session}, and one \textit{screen} session contains at least one \textit{app} session) and stayed on the app for longer than 5 seconds (we set this threshold to avoid asking participants too frequently when they fast switch between apps), the data-collection app would display a survey asking why they opened the app. When designing this ESM survey, we adopted the Uses and Gratifications (U\&G) types in~\cite{lukoff2018} and slightly modified the original categories by combining habitual use into the U\&G types, which we called ``no specific goal.'' This was mainly to reduce participants' effort of answering two questions. Our survey question prompted the user to report their intention by asking ``why are you here'' and had 7 options: ``no specific goal,'' ``get things done or self-improve,'' ``get information,'' ``communicate or interact with others,'' ``be entertained or have fun,'' ``browse social media,'' and ``I am not sure.'' See Table~\ref{tab:u&g} for the full definitions of these categories, and the abbreviations we use for the rest of the paper. The ESM prompt interface is shown in Figure~\ref{fig:app}. After the participant answered this question, the data-collection tool would start taking screenshots in the background every 5 seconds. To implement this, we used the \texttt{MediaProjection} API~\cite{android2024mediaprojection} on Android, which allowed our app to record screen content.

At 8 PM everyday, the app sent a notification to the participant to remind them to answer questionnaires about how regretful they felt about the app sessions captured by the data-collection tool. For each session, the app presented to them the starting time, the app name, the screenshots captured in the session, followed by a survey question \revise{measuring their regret for the session. Although there are existing standardized scales for measuring regret, such as the five-item Decision Regret Scale~\cite{decision_regret_scale}, we opted to use a simpler, one-item survey question, another common approach used in past studies to measure regret (e.g.,~\cite{van2008, sagi2007, allaert2019}). This is because we needed participants to rate every app session, and we hoped that by using a lightweight survey question we would avoid overburdening participants and prevent a decline in response quality due to lengthy surveys~\cite{galesic-length-quality}. The instruction and wording of our survey question resembles those used in~\cite{decision_regret_scale, van2008} and were adapted to the context of phone use. Also, given that regret is a rich and complex emotion (even when narrowed down only to action-based regret) that prior work defines broadly~\cite{gilovich1995}, we imposed minimal constraint in our survey question. Specifically, we asked:} ``Please think about this particular time you used your phone. Select how you feel about the following statement: I feel regret about this phone use session.'' Participants answered this question on a seven-point Likert scale, from ``Strongly Disagree,'' to ``Strongly Agree.'' All of the interfaces are shown in Figure~\ref{fig:app}. \revise{One reason for assessing participants' regret at the end of each day at 8 PM (when people are generally more available), not immediately after each session, is that we wanted to avoid disrupting their normal phone use. More importantly, since regret is inherently a cognitive-laden emotion~\cite{gilovich1995} which involves comparing alternative options~\cite{zeelenberg2007}, participants may need some time to reflect on how much they regret a phone use session, which makes in-the-moment assessment potentially less reliable.}

\renewcommand{\arraystretch}{1.35}
\begin{table*}
  \caption{``Intended Use'' categories shown in the ESM prompt and their definitions, based on Lukoff et al.'s application of Uses \& Gratifications Theory~\cite{lukoff2018}.}
  \label{tab:u&g}
  \Description{
  The table consists of three columns: Category, Abbreviation, and Definition. It outlines different intentions for phone use and the associated definitions.
  }
  \begin{tabular}{ c c m{7cm} }
  \toprule
    Category & Abbreviation & \multicolumn{1}{c}{Definition} \\
    \hline
    No specific goal & \textit{No Specific Goal} & You use your phone habitually or without a clear goal, to browse, explore, or pass the time. \\
    \hline
    Get things done or self-improve & \textit{Productivity} & You want to achieve specific tasks and engage in activities focused on productivity or personal development. \\
    \hline
    Get information & \textit{Information} & You want to acquire knowledge or stay updated on various topics. \\
    \hline
    Communicate or interact with others & \textit{Communication} & You want to connect and interact with people through messaging, social networking, or video calls. \\
    \hline
    Be entertained or have fun & \textit{Entertainment} & You want to relax and enjoy activities such as watching movies, listening to music, or playing video games. \\
    \hline
    Browse social media & \textit{Social} & You want to consume content on social media without actively engaging, such as scrolling on feeds. \\
    \hline
    I am not sure & N/A (excluded) & You do not know how your motivation for phone use fits into these categories, you are using your phone accidentally, or you do not have time to answer this survey question. \\
    \bottomrule
  \end{tabular}
\end{table*}

\subsubsection{Two Follow-up Interviews} On two of the seven days during the week of data collection, participants engaged in a one-hour interview with a member of the research team. We scheduled the two interviews such that each participant could reflect on their phone use both on a weekday and on the weekend\revise{, as their behavior might differ between weekdays and weekends due to different daily routines}. During the interview, the participants used screen sharing to go through all the sessions captured by the app from that day or the previous day, and were asked to reflect on why they regretted some experiences and valued others. \revise{As participants went through each session, the interviewer asked questions to understand what led to regret (or satisfaction), such as: ``\textit{Can you think out loud to help me understand exactly how you feel about the session and the reason why you felt (or did not feel) regret about the session?}'' The interviewer also asked about participants' intention for app use when reviewing some sessions, such as: ``\textit{What motivated you to use this app at the time?}'' At the end of each follow-up interview, the interviewer asked participants to reflect holistically about what experience they found regrettable. For example, we asked: ``\textit{Can you summarize how you answered the regret survey question? What factors did you consider when determining whether a situation is regretful or not?}''} All of the follow-up interviews were conducted via Zoom.

\subsection{Data Analysis}
\subsubsection{Interview data} Since our initial interviews were brief orientations, we focused our qualitative data analysis on the 34 follow-up interviews (two follow-up interviews for each participant). To conduct the qualitative data analysis, we used the collaborative qualitative data analysis tool Dovetail~\footnote{https://dovetail.com/role/researcher/}\revise{, which has been used in past HCI studies~\cite{ko2023dovetail, tu2024dovetail, boyd2023dovetail} and is HIPAA, GDPR, and CCPA compliant~\cite{dovetail_trust}, indicating robust privacy protection. All of the interview data uploaded to Dovetail was anonymized.} We first used the automated transcription service in Dovetail to turn the 34 audio recordings into text. Then following the thematic analysis approach~\cite{thematic}, two researchers first separately reviewed the transcripts and came up with initial codes \revise{in an inductive-deductive manner, generating codes both based on the data and on our research questions}. Then the two researchers met and collaboratively constructed codes and applied them to all the transcripts. \revise{Our codebook is shared in Appendix~\ref{codebook}.} After coding the data, one researcher pulled out important quotes from the transcripts. For each quote that appears in the paper, we use ``PX, FX'' to denote the participant ID and interview ID (1 being the first follow-up interview, and 2 being the second).

\subsubsection{ESM and survey data} We collected data from 1,631 screen sessions and 3,946 app sessions in total, which add up to approximately 183 hours of phone use. For each app session, we had the associated app name, starting time, duration, intended use (Table~\ref{tab:u&g}), and regret (in a seven-point Likert scale). For our quantitative data analysis, we used analysis of variance based on mixed ordinal logistic regression, as our response variable (regret) is an ordinal variable.

\subsubsection{Screenshot data}\label{mtd-screenshot} We collected 119,373 screenshots in total. We focus our screenshot data analysis on social media apps, since social media usage is one of the most common cases of problematic smartphone use~\cite{lukoff2018}, and these apps have mixed features such as recommendation, search, direct messaging, requiring a nuanced understanding~\cite{cho2021}, which screenshots can provide. We chose social media apps that had been used more than 50 times and at least by 3 participants from our dataset. We also excluded apps that do not have a ``feed'' feature and only support direct messaging (e.g., Messenger, WhatsApp). Initially, we obtained a list of 6 apps including: \textit{Instagram}, \textit{Facebook}, \textit{X}, \textit{Snapchat}, \textit{TikTok}, and \textit{Reddit} (in the order of their occurrences in our dataset from high to low)\revise{, totaling 38,017 screenshots}.

We constructed seven categories of social media activity based on common behavior of social media users, and we considered both user action and the source of content, an approach used in prior work~\cite{cho2021, orzikulova2023}. The seven categories include: \textbf{Active Communication}, \textbf{Active Search}, \textbf{Consuming Recommendation-Based Content}, \textbf{Consuming Subscription-Based Content} (this includes when the user is viewing content from accounts they follow), \textbf{Consuming Content Shared by Others}, \textbf{Viewing Comments or Discussion Thread}, and \textbf{None of the Above} (See Tables~\ref{tab:activity-def-1} and~\ref{tab:activity-def-2} for detailed definitions of these categories and example screenshots).

\begin{table*}
  \caption{Social media activity categories, the corresponding abbreviation used in the rest of the paper, definition used for human and AI coding, and one example screenshot (all private messages and identifiable information have been redacted).}
  \label{tab:activity-def-1}
  \Description{
  The table consists of four columns: Category, Abbreviation, Definition, and Example Screenshot. It outlines different social media activity types, their definitions, and representative screenshots. For Active Communication (Communication), the example screenshot shows a chat interface with a keyboard. For Active Search (Search), the example screenshot displays a TikTok interface with search results. For Consuming Recommendation-Based Content (View_Recommendation), the example screenshot shows a recommended content feed with suggested posts. For Consuming Subscription-Based Content (View_Subscription), the example screenshot shows a social media feed with content from subscribed sources (nytimes).
  }
  \begin{tabular}{m{3cm} C{3cm} m{7.5cm} C{3cm}}
    \toprule
    Category & Abbreviation & \multicolumn{1}{c}{Definition} & Example Screenshot\\
    
   \midrule
   Active Communication & \textit{Communication} & The screenshot includes the presence of a private messaging interface, suggesting the user is actively communicating with specific individuals or groups. &  \includegraphics[scale=0.06]{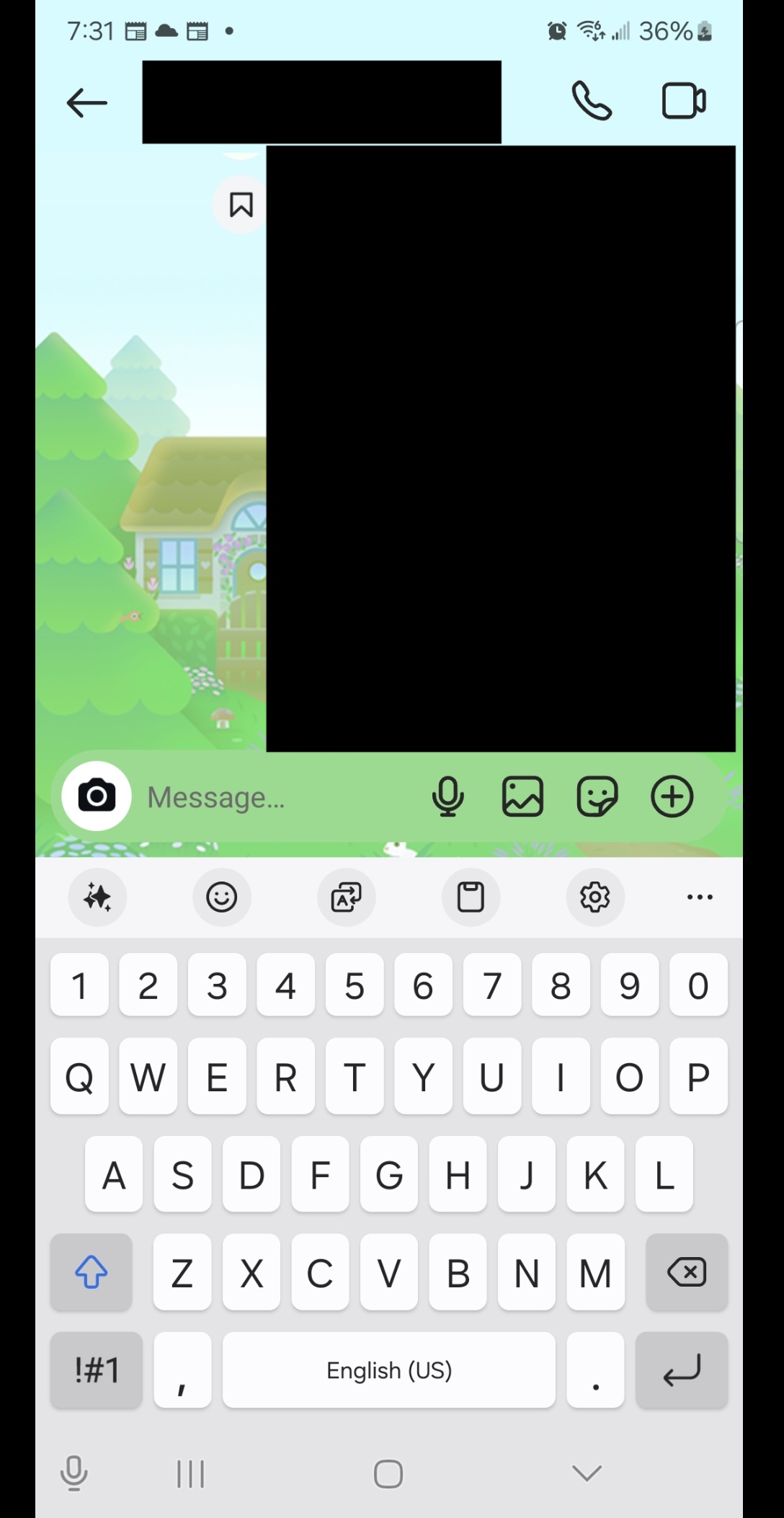} \\
    \midrule
    Active Search & \textit{Search} & The screenshot suggests the user is using the search feature to find specific information, content, articles, or items or is consuming content they found through active searching. & \includegraphics[scale=0.06]{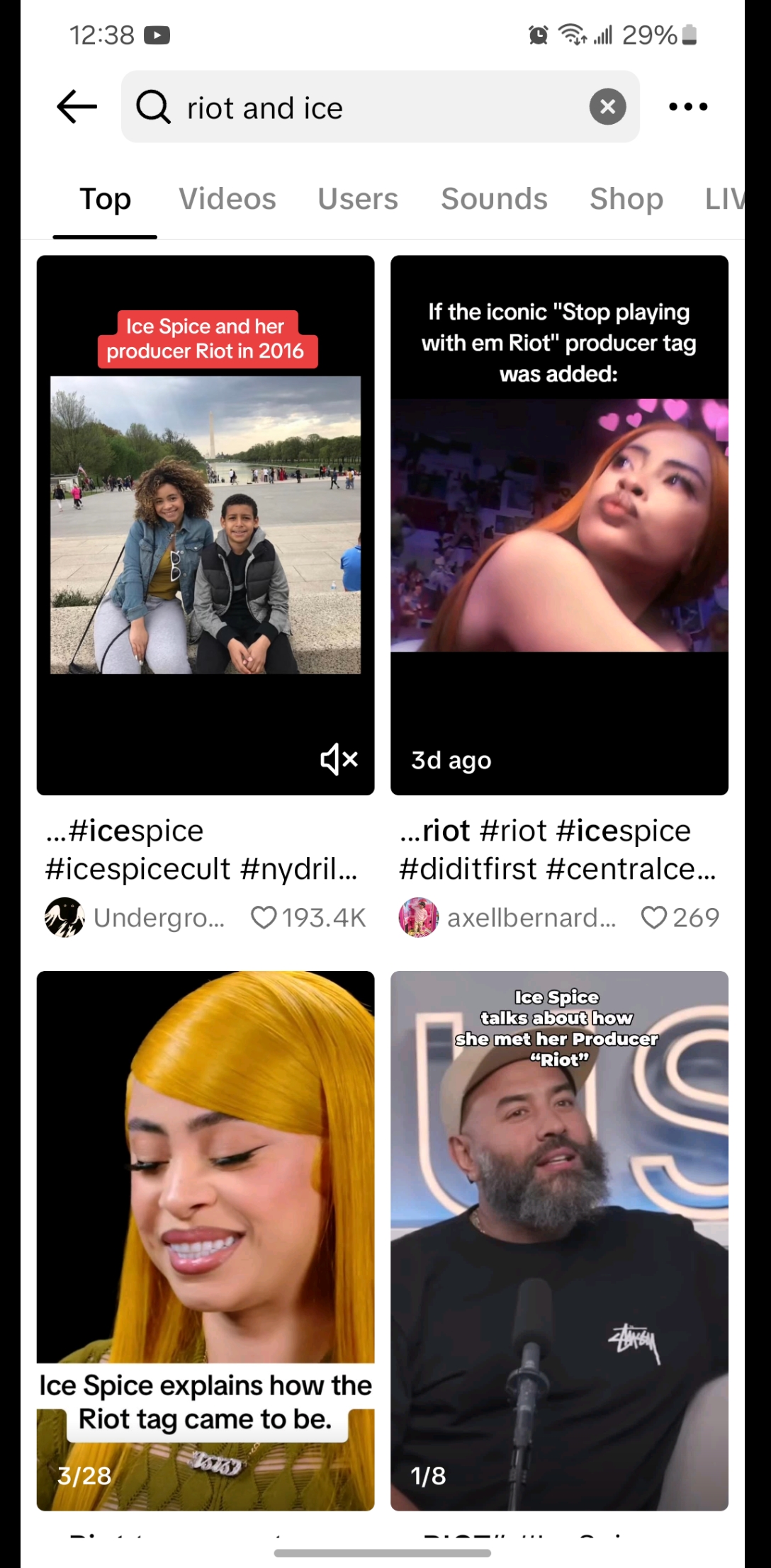} \\
    \midrule
    Consuming Recommendation-Based Content & \textit{View\_Recommendation} & The screenshot shows explicit indicators of recommendation-based content in the user's feed, such as a ``For You'' tab, ``Suggested Post'' labels, or buttons like ``Follow'' and ``Join'' which allow users to subscribe to new content. & \includegraphics[scale=0.06]{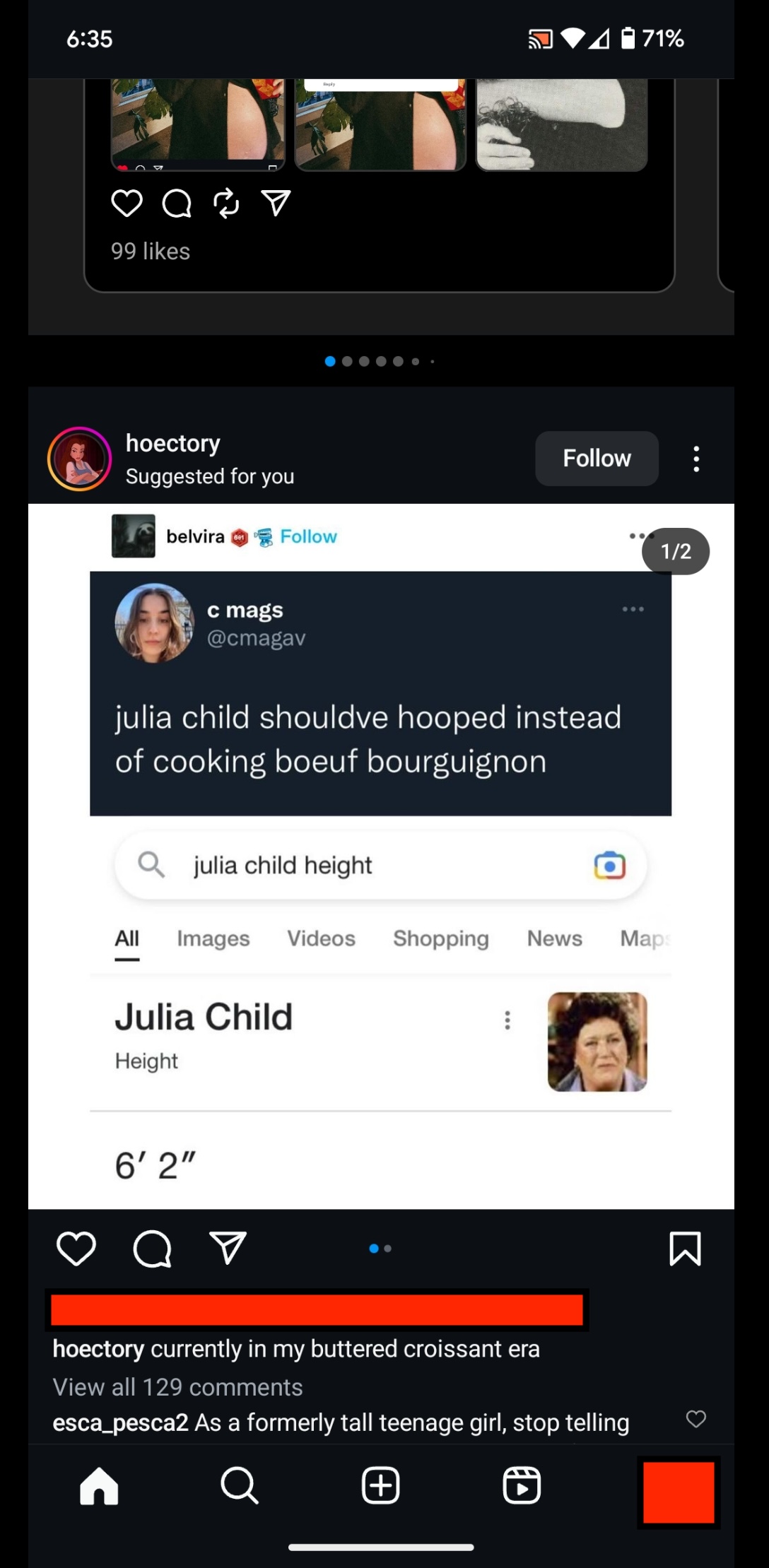}\\
    \midrule
    Consuming Subscription-Based Content & \textit{View\_Subscription} & The screenshot shows content already followed or subscribed to by the user in their feed. The screenshot should include indicators such as an active ``Following'' or ``Subscription'' tab of the app, or signs suggesting that the user has already followed the content poster, such as the absence of buttons next to the content poster or community to follow, join, or subscribe in a feed interface. & \includegraphics[scale=0.06]{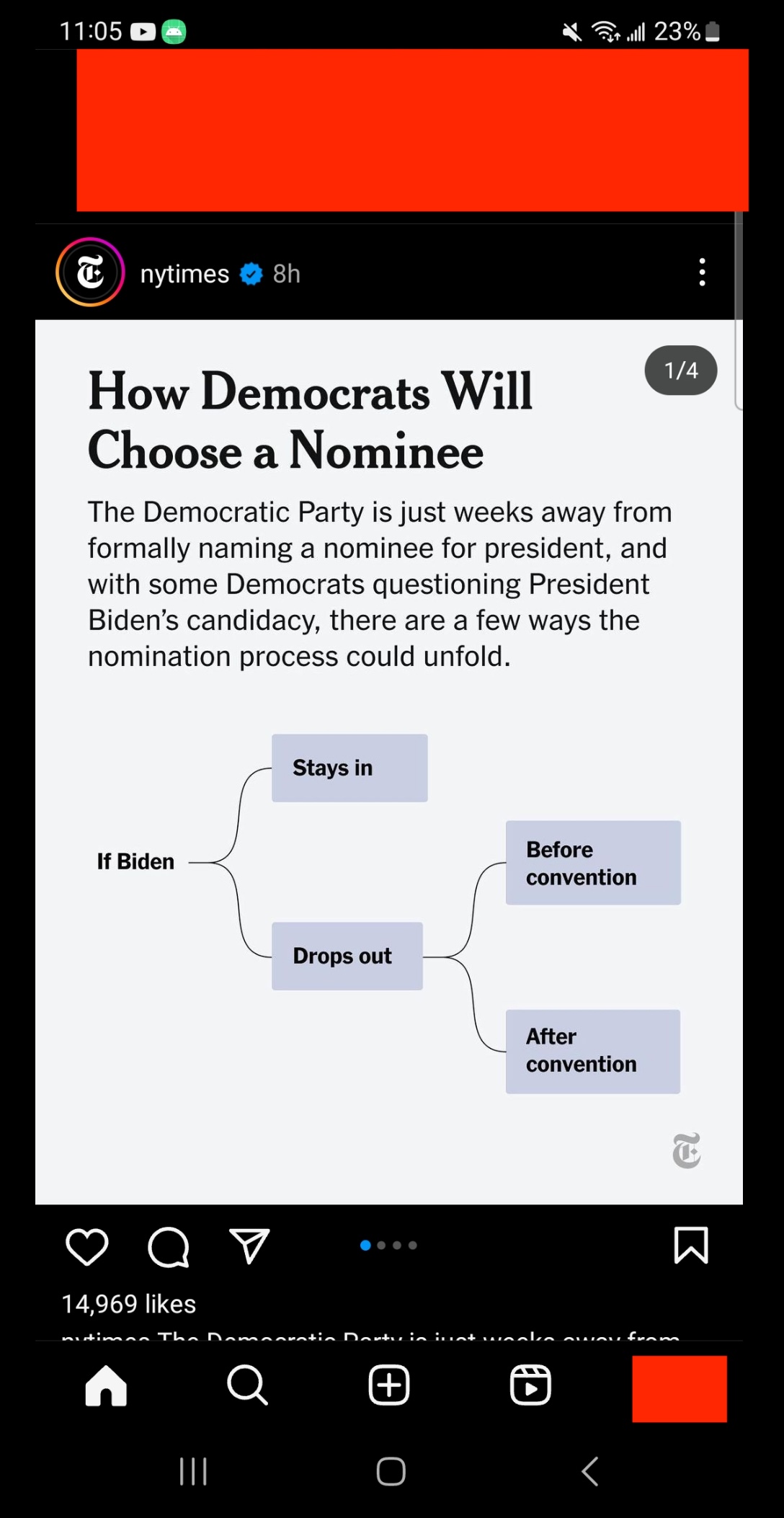}\\
    
    \bottomrule
  \end{tabular}
\end{table*}

\begin{table*}
  \caption{(Continued) Social media activity categories, the corresponding abbreviation used in the rest of the paper, definition used for human and AI coding, and one example screenshot (all private messages and identifiable information have been redacted).}
  \label{tab:activity-def-2}
  \Description{
  The table consists of four columns: Category, Abbreviation, Definition, and Example Screenshot. It categorizes different types of social media activity, providing definitions and example screenshots. For Consuming Content Shared by Others (View_Shared), the example screenshot displays a shared post (a friend of the user shared a post created by someone the user does not follow). For Viewing Comments or Discussion Thread (View_Comments), the example screenshot shows a comment thread under a post. For None of the Above (Other), the example screenshot shows a T-Mobile sponsored ad on TikTok.
  }
  \begin{tabular}{m{3cm} C{3cm} m{7.5cm} C{3cm}}
    \toprule
    Category & Abbreviation & \multicolumn{1}{c}{Definition} & Example Screenshot\\
    
   \midrule
    Consuming Content Shared by Others & \textit{View\_Shared} & The screenshot suggests the user is viewing content shared or reposted by someone they followed or opened from a private conversation. & \includegraphics[scale=0.085]{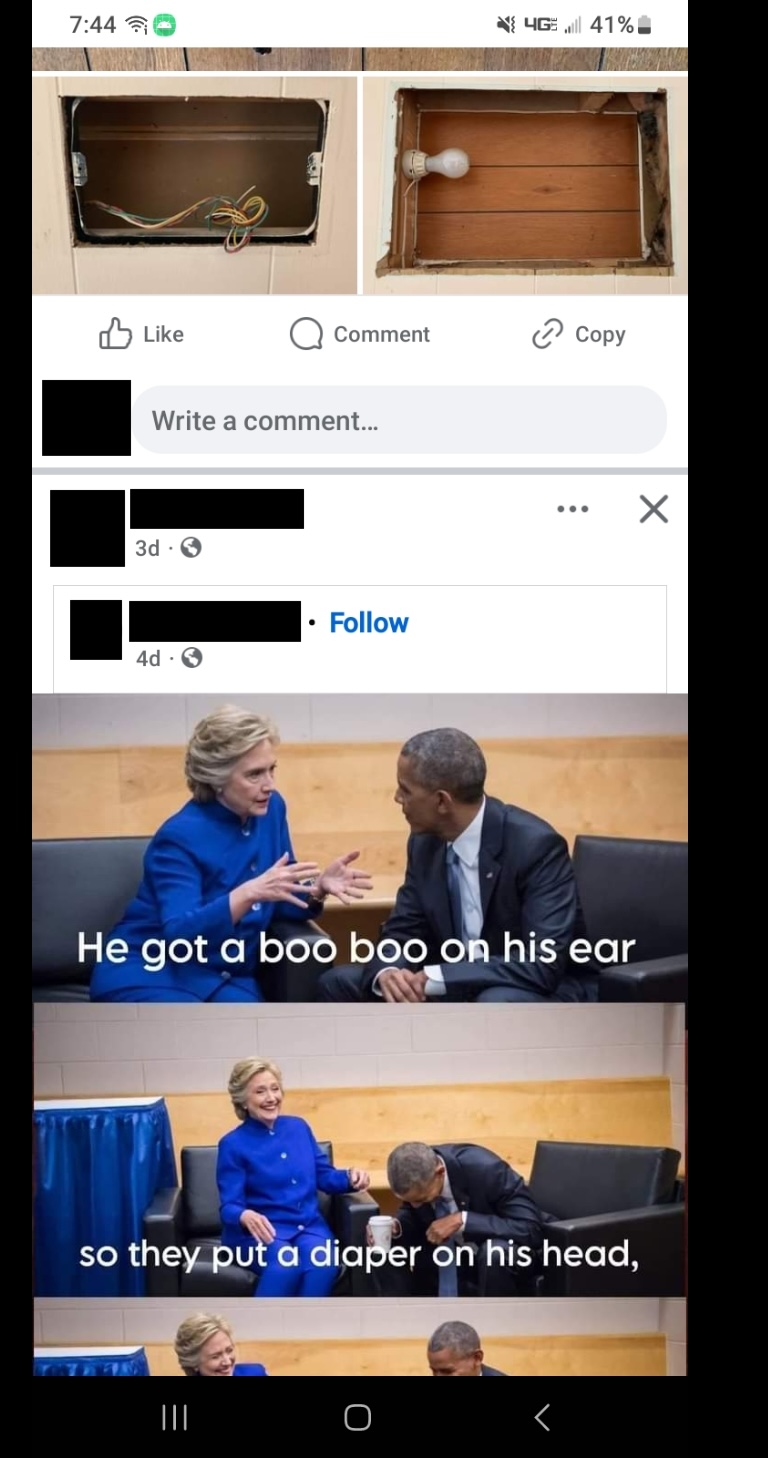}\\
    \midrule
    Viewing Comments or Discussion Thread & \textit{View\_Comments} & The screenshot suggests the user is viewing the comment section or discussion thread of a social media post. & \includegraphics[scale=0.085]{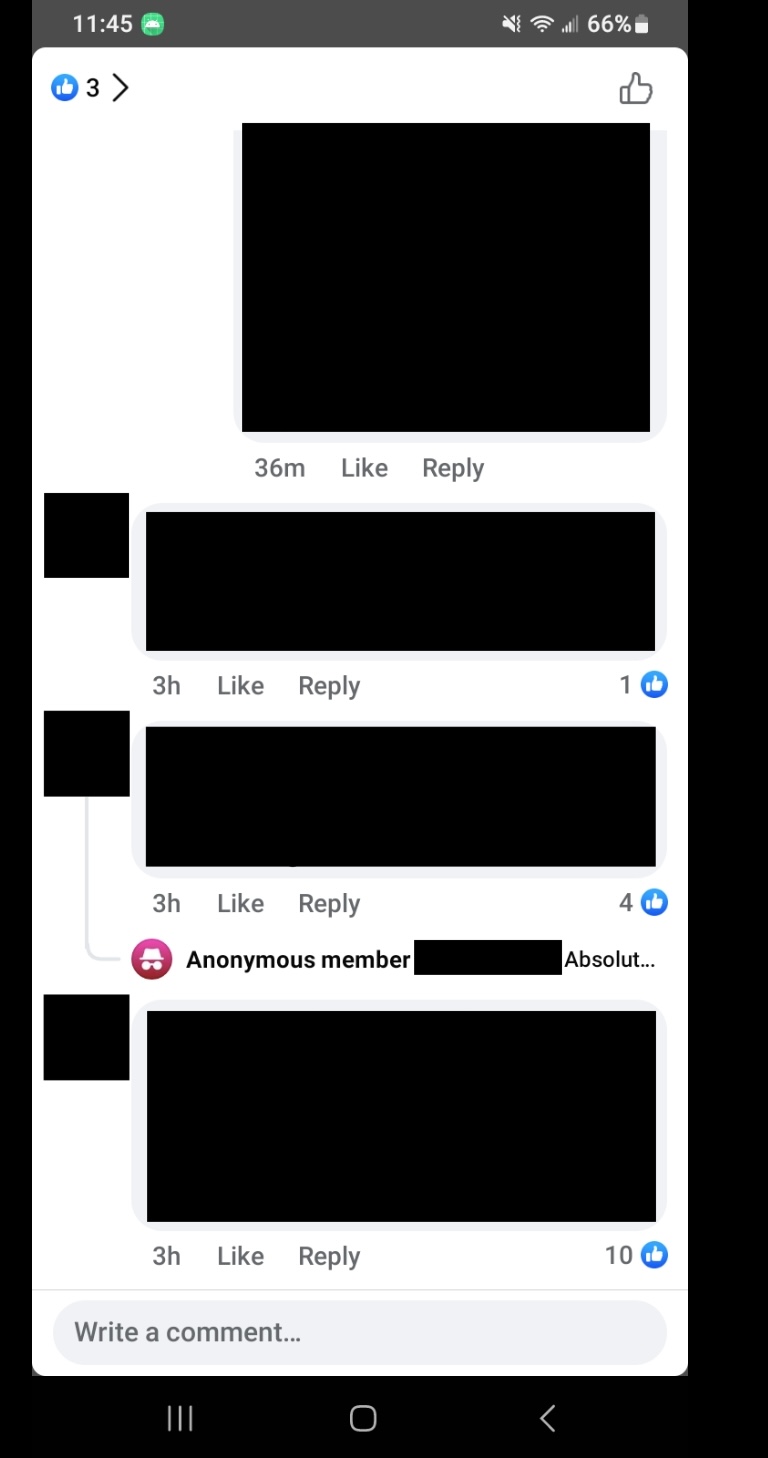} \\
    \midrule
    None of the Above & \textit{Other} & When the user opens a link to a website, when the user sees sponsored content or ad, when it is unclear if user is viewing content posted by someone they followed or recommended to them, or when seeing these screens: home screen, notification screen, a black, dimmed, or blank screen, a screen showing a survey prompt. & \includegraphics[scale=0.06]{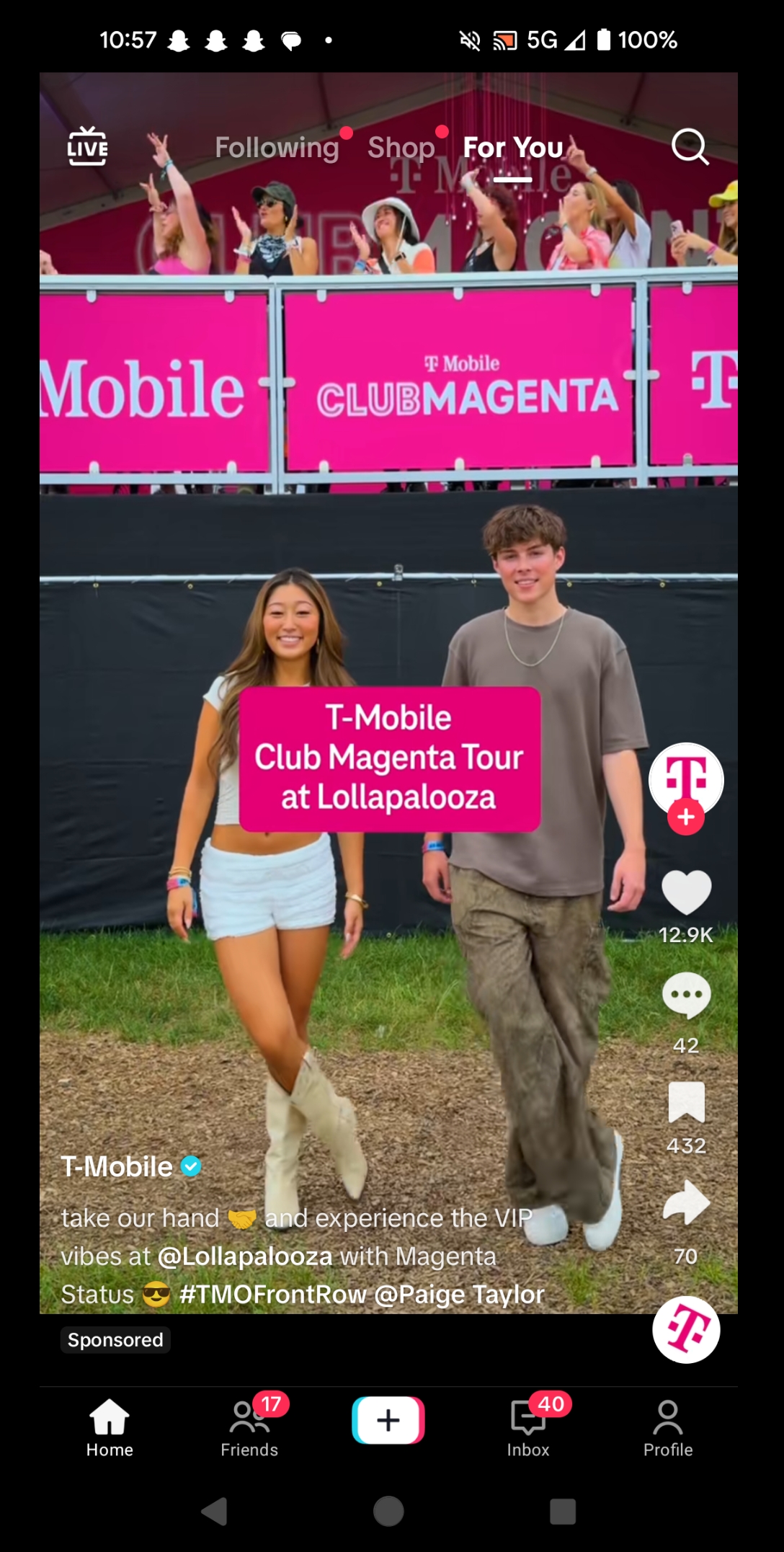} \\
    
    \bottomrule
  \end{tabular}
\end{table*}

\begin{figure}
    \centering
    \includegraphics[width=\linewidth]{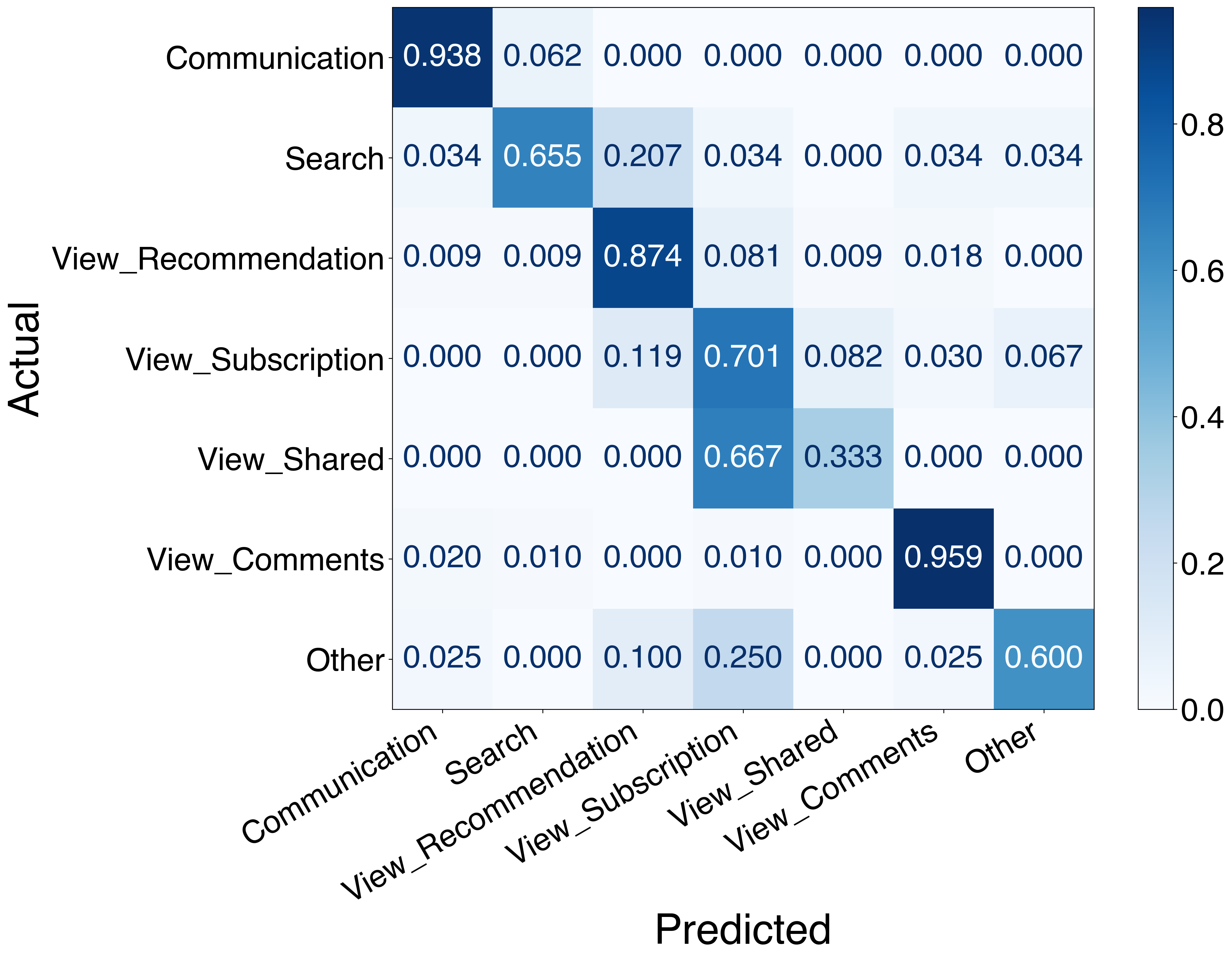}
    \caption{\revise{Normalized confusion matrix. Each row represents one actual class, and each column represents one predicted class. The number in each cell represents the proportion of a predicted class for a given actual class.}}
    \label{fig:confusion-matrix}
    \Description{
    A confusion matrix heatmap comparing actual versus predicted classifications for different categories, including “Communication”, “Search”, “View_Recommendation”, “View_Subscription”, “View_Shared”, “View_Comments”, and “Other”. The diagonal elements show the highest values, indicating correct classifications, while off-diagonal elements represent misclassifications. The intensity of the color corresponds to classification accuracy, with darker blue indicating higher values. The x-axis represents predicted labels, and the y-axis represents actual labels. A color bar on the right provides a reference for the scale of values. “Communication” was correctly classified 93.8\% of the time but misclassified as “Search” 6.2\% of the time. “Search” was correctly classified 65.5\% of the time, with 20.7\% misclassified as “View_Recommendation”. “View_Recommendation” was correctly classified 87.4\% of the time but misclassified as “View_Subscription” (8.1\%) and “View_Comments” (1.8\%). “View_Subscription” had a 70.1\% accuracy, with 11.9\% misclassified as “View_Recommendation” and 8.2\% as “View_Shared”. “View_Shared” was misclassified 66.7\% of the time as “View_Subscription”, with a correct classification rate of 33.3\%. “View_Comments” had the highest accuracy at 95.9\%, with minimal misclassifications. “Other” was correctly classified 60.0\% of the time, with notable misclassification into “View_Recommendation” (10.0\%) and “View_Subscription” (25.0\%).
    }
\end{figure}

We used OpenAI's multimodal large language model, \texttt{gpt-4o}, which showed exceptional ability in visual tasks~\cite{openai2024gpt4o}, to help code the screenshots. Before applying screenshot analysis to the entire dataset using large language model, we assessed the consistency between human coders and \texttt{gpt-4o} to assess if the model could successfully categorize the screenshots based on our coding scheme. We randomly picked 500 screenshots from the dataset  and let two researchers categorize them (we also included the previous four screenshots as context). The interrater reliability between the two human raters, measured in Cohen's Kappa~\cite{cohen1960kappa}, was 0.82 (with a percentage agreement of 86\%), indicating almost perfect agreement. The two raters then discussed to resolve conflicts, which led to ``human consensus'' categories, which we then compared with categories generated by \texttt{gpt-4o} using text descriptions of the screenshots with zero-shot and chain-of-thought prompting~\cite{zeroshotllm} (detailed process below). The interrater reliability between \texttt{gpt-4o} and human consensus is 0.72 (with a percentage agreement of 78\%), indicating substantial agreement. When breaking our results down by app (see Table~\ref{tab:reliability}), we found that X had the lowest human-human consistency and the lowest raw percentage consistency among all the apps. This aligned with our two raters' observation. We found that on X, the recommendation feed (``For You'') and the following feed (``Following'') often look identical when the top bar is hidden, making it challenging to detect whether the user is viewing recommendation-based or subscription-based content. Since we cannot reliably detect the difference between these two activities on X, we dropped X for our data analysis. With the rest of the apps, our classification showed a human-AI consistency of 0.74 measured in Cohen's Kappa and 79\% measured in raw percentage agreement, indicating substantial agreement. \revise{To further assess the validity of LLM labeling, we plotted the confusion matrix (see Figure~\ref{fig:confusion-matrix}) and examined the precision, recall, and F1 score for each class (Table~\ref{tab:class-metric}). Results showed that the weighted average of F1 scores achieved nearly 0.8, with most of the classes showing an F1 score larger than 0.7 (including \textit{Communication}, \textit{Search}, \textit{View\_Recommendation}, \textit{View\_Subscription} and \textit{View\_Comments}). However, the precision, recall, and F1 score for the class \textit{View\_Shared} are particularly low (0.2). Out of its 6 instances that appeared in the test set, 4 were misclassified as \textit{View\_Subscription}. 11 instances of \textit{View\_Subscription} were misclassified as \textit{View\_Shared}. This might be due to the fact that, it is easy to confuse content \textit{shared} by a subscribed account with content \textit{posted} by a subscribed account (refer to Tables~\ref{tab:activity-def-1} and~\ref{tab:activity-def-2} for a comparison). Since this category is rare (accounting for only 1\% of the data in the test set), it likely will have a minimal impact on our analysis. We still keep this category in the analysis, but caution readers that the data associated with this small category are likely erroneous.} Our final social media use dataset includes 664 app sessions and 34,313 screenshots, adding up to approximately 50 hours of social media use. 

\begin{table}
  \caption{Human-human and human-AI interrater reliability by app based on Cohen's Kappa and raw percentage agreement. X (formerly Twitter) was later dropped due to having the lowest human-human consistency caused by ambiguity of the interface. The row in bold represents reliability for the remaining apps.}
  \label{tab:reliability}
  \Description{
  The table presents interrater reliability metrics for different apps, comparing human-human consistency and human-AI consistency using Cohen’s Kappa and raw percentage agreement. The table consists of 4 columns on the top: 1.	App: The social media platform analyzed. 2. \% in the Test Set: The proportion of the dataset corresponding to each app. 3. Human-Human Consistency. 4. Human-AI Consistency. For both of the consistencies, Cohen’s Kappa and Raw percentage agreement are reported. Key stats are reported in the text. 
  }
  \begin{tabular}{c C{1cm} C{1cm} C{1cm} C{1cm} C{1cm}}
    \toprule
    \multirow{3}{*}{App} & \multirow{-0.5}{1cm}{\% in the Test Set} & \multicolumn{2}{C{2.5cm}}{Human-Human Consistency \vspace{1mm}} & \multicolumn{2}{C{2.5cm}}{Human-AI Consistency \vspace{1mm}} \\
    \cline{3-6}

    & & \vspace{1mm} Cohen's Kappa & \vspace{1mm} Raw \% & \vspace{1mm} Cohen's Kappa & \vspace{1mm} Raw \% \\
    
    \midrule
    All & 100.0\% & .816 & 85.6\% & .715 & 77.6\%  \\
    \textbf{All (-X)} & \textbf{86.8\%} & \textbf{.842} & \textbf{87.8\%} & \textbf{.736} & \textbf{79.5\%}  \\
    Facebook & 33.8\% & .845 & 88.8\% & .664 & 75.1\%  \\
    Instagram & 31.6\% & .814 & 86.7\% & .757 & 82.3\%  \\
    TikTok & 15.6\% & .792 & 85.9\% & .749 & 83.3\%  \\
    X & 13.2\% & .648 & 71.2\% & .575 & 65.2\%  \\
    Reddit & 4.8\% & .830 & 91.7\% & .528 & 79.2\%  \\
    Snapchat & 1.0\% & 1.000 & 100.0\% & .545 & 80.0\%  \\
    \bottomrule
  \end{tabular}
\end{table}

\begin{table}
  \caption{\revise{Evaluation metrics for each class (representing one type of social media activity), including precision, recall, and f1 score. ``Support'' indicates the number of actual occurrences of the class in the test set. Micro (accuracy), macro, and weighted averages are also reported.}}
  \label{tab:class-metric}
  \Description{
  This table presents the performance metrics of a classification model for different social media activity classes. It includes precision, recall, F1 score, and support (the number of actual occurrences of each class in the test set). 	Accuracy: .795. Macro Average: .709 Precision, .723 Recall, .706 F1 Score. Weighted Average: .806 Precision, .795 Recall, .797 F1 Score. F1 scores for each class: Communication: 0.833, Search: 0.745, View_Recommendation: 0.829, View_Subscription: 0.743, View_Shared: 0.200, View_Comments: 0.940, Other: 0.649.
  }
  \begin{tabular}{c c c c c}
    \toprule
    Class & Precision & Recall & F1 Score & Support \\
    \midrule
    \textit{Communication} & .750 & .938 & .833 & 16 \\
    \textit{Search} & .864 & .655 & .745 & 29 \\
    \textit{View\_Recommendation} & .789 & .874 & .829 & 111 \\
    \textit{View\_Subscription} & .790 & .701 & .743 & 134 \\
    \textit{View\_Shared} & .143 & .333 & .200 & 6 \\
    \textit{View\_Comments} & .922 & .959 & .940 & 98 \\
    \textit{Other} & .706 & .600 & .649 & 40 \\
    Accuracy & .795 & .795 & .795 & 434 \\
    Macro Average & .709 & .723 & .706 & 434 \\
    Weighted Average & .806 & .795 & .797 & 434 \\
    \bottomrule
  \end{tabular}
\end{table}

\begin{figure*}
    \centering
    \includegraphics[width=\linewidth]{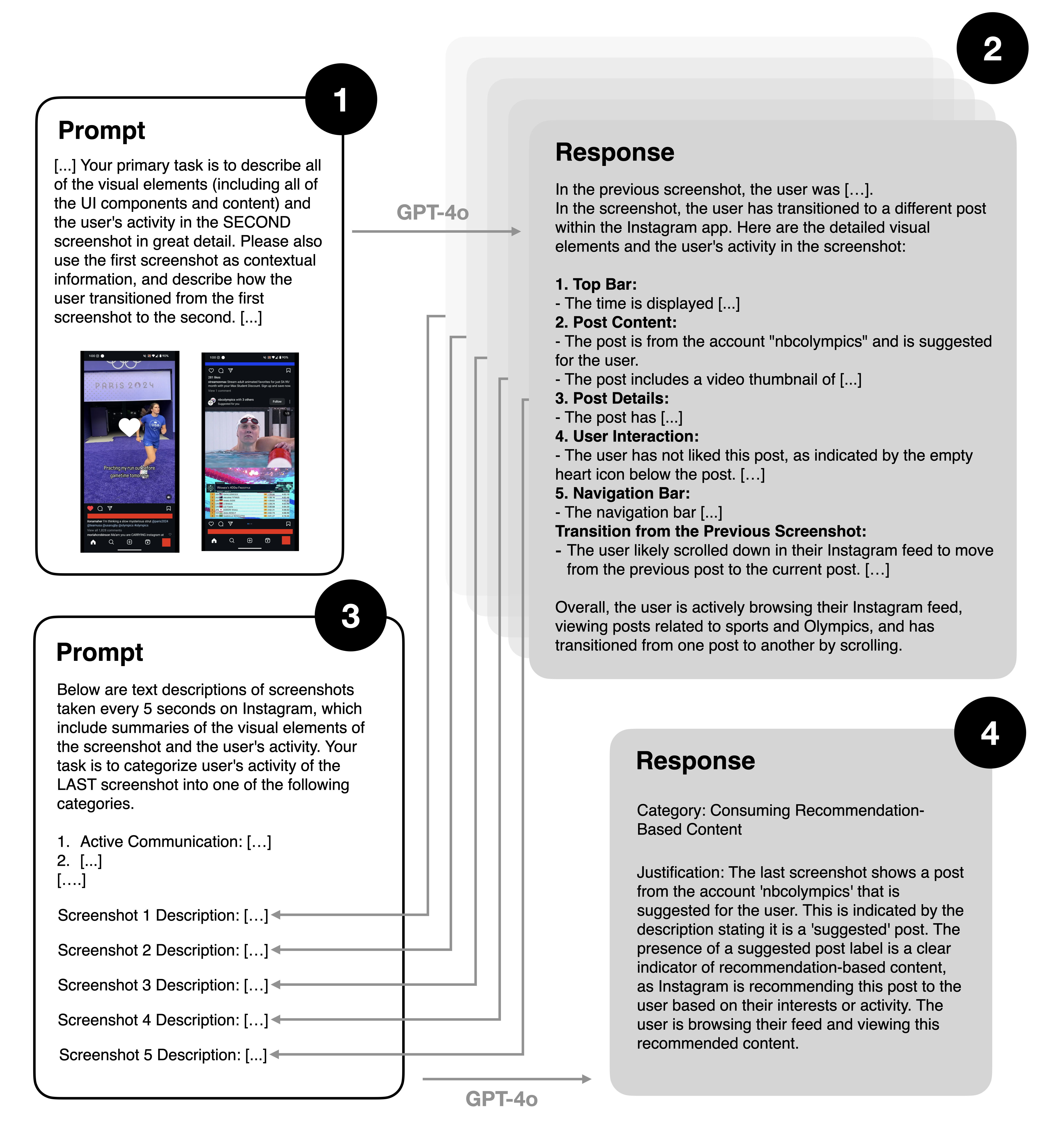}
    \caption{The process of using GPT-4o to code screenshots. The associated activity category for each screenshot was obtained using the steps shown in the diagram. (1) An image-to-text prompt was first constructed, which included instructions to describe all visual elements on the screenshot and how the user transitioned from the previous screenshot in the same session (if one exists). (2) The model sent back detailed text description of the screenshot. (3) Combining the text descriptions of the previous four screenshots, we constructed the second text-only prompt which asked the model to categorize the user's activity. (4) The model sent back the category it identified and its justification for choosing that category.}
    \label{fig:llm}
    \Description{
    A visual diagram illustrating a multi-step process for analyzing smartphone screenshots using GPT-4o. The diagram consists of four numbered sections: 1.	Prompt (Step 1): A request for GPT-4o to describe all visual elements and user activities in the second screenshot, using the first screenshot as contextual information. The prompt includes two example Instagram screenshots. 2.	Response (Step 2): GPT-4o’s output, providing a structured breakdown of the second screenshot, including elements like the top bar, post content, post details, user interaction, and navigation bar. It also describes how the user transitioned from the first to the second screenshot. 3.	Prompt (Step 3): A request for GPT-4o to categorize user activity in the last of a sequence of five screenshots based on predefined categories. The prompt provides textual descriptions of each screenshot. 4.	Response (Step 4): GPT-4o’s classification of the user’s activity in the last screenshot as “Consuming Recommendation-Based Content,” with a justification stating that the post was labeled as “suggested,” indicating recommendation-based browsing.
    }
\end{figure*}

To automatically code these screenshots, we first turned all screenshots into textual descriptions using \texttt{gpt-4o}, as we observed that reasoning directly on raw images could result in poorer performance compared to reasoning on textual descriptions of screenshots, particularly when multiple screenshots are involved. We asked the model to ``\textit{describe all of the visual elements (including all of the UI components and content) and the user's activity in the screenshot in great detail}.'' Importantly, when asking the model to generate descriptions for each screenshot, we also provided the previous screenshot in the same app session (if applicable) as context and asked the model to describe ``\textit{how the user transitioned from the first screenshot to the second}'' to help the model identify transitions and relationships between screenshots. The full prompt is included in Appendix~\ref{prompt1}. All screenshots were sent to \texttt{gpt-4o} with 1/16 of the original resolution (1/4 of the original width and height respectively), which provided a good balance of cost/speed and fidelity.

Then, using the textual descriptions generated by \texttt{gpt-4o}, we asked the model to categorize each screenshot into one of seven categories of social media activity using structured outputs~\cite{openai2024structuredoutputs}. We employed zero-shot prompting with chain-of-thought reasoning, which have been shown to elicit multi-step reasoning and help reach correct answers~\cite{zeroshotllm}. When providing the textual descriptions, we also included the descriptions of the previous four screenshots in the same app session (if applicable) as context, since some critical details used for categorization may be hidden when looking at only one screenshot. Note that the four previous screenshots were also provided when human coders were coding the screenshots used for assessing consistency. We provided the definition of each of the social media categories to the model, including visible indicators of the categories (such as a highlighted ``For You'' tab for Consuming Recommendation-Based Content), similar to how a human coder would classify these screenshots. We then complemented this with a few general rules, such as asking the model to pay attention to the previous screenshots to infer user's current activity. The full prompt is included in Appendix~\ref{prompt2}. See Figure~\ref{fig:llm} for a graphical example of the MLLM coding process.

\subsection{Ethical Considerations}
We carefully considered how to protect participants' privacy, given the invasive and potentially sensitive nature of our data collection. We implemented several mechanisms to give participants control over their data. First, we ensured that screenshots were only stored locally and shared with the research team only when the participant actively elected to upload them. \revise{All of the uploaded screenshots were then stored in a private Amazon S3~\footnote{https://docs.aws.amazon.com/s3/} bucket, which automatically applies server-side encryption~\cite{aws_s3_encryption} and can only be accessed by using the administrator credentials of the bucket owner.} Second, participants could view all of the screenshots captured by the app and delete an entire session of screenshots at any time. They were also asked to review and delete any screenshots they did not want us to see when reflecting on the data and filling in daily questionnaires at the end of each day. Third, if a participant anticipated that they would be doing something private on their phone (such as checking their bank account), they were allowed to temporarily stop data collection and resume participating later. 

During the initial interview and before the participants explicitly consented to join the study, we were fully transparent about when and how passive screenshots would be taken during the study, and how the privacy protection mechanisms mentioned above work. We also explained to participants that only researchers in the team will be able to access the raw screenshot data, and that we may use the screenshots to train our own machine learning models and use OpenAI's GPT models to analyze the screenshots. We cited OpenAI's policy which states that they do not store image data uploaded via their API or use the data for training their model~\cite{openai2024visionguide}. \revise{To ensure that the participants fully understood how the screenshot data would be collected and used, we also asked them if they needed any clarification or had any concerns regarding data privacy before having them join the study.} \revise{Since our study required participants to constantly review and reflect on their phone use behavior, which can cause emotional stress, we emphasized that their participation was voluntary, and that they could drop out of the study at any time if they did not feel comfortable, in which case they would still receive pro-rated compensation for any portion of the study they had completed.} \revise{In terms of data sharing with OpenAI, we eventually used 34,312 raw screenshots encoded in the base64 format in our requests, along with the associated app name. No other information about the participants was shared with OpenAI.} This study was reviewed by our institutional review board (IRB) and deemed exempt.

\section{Results}

\subsection{How Regret Varies by User Intention}
\revise{To address RQ1, w}e began by analyzing all 3,946 sampled app sessions \revise{of general phone use} to explore the relationship between regret and user intention. These sessions included any app usage, except for Settings and the data-collection app itself.

\subsubsection{Regret varies based on intended use}\label{regret-intention}
\begin{figure}
    \centering
    \includegraphics[width=\linewidth]{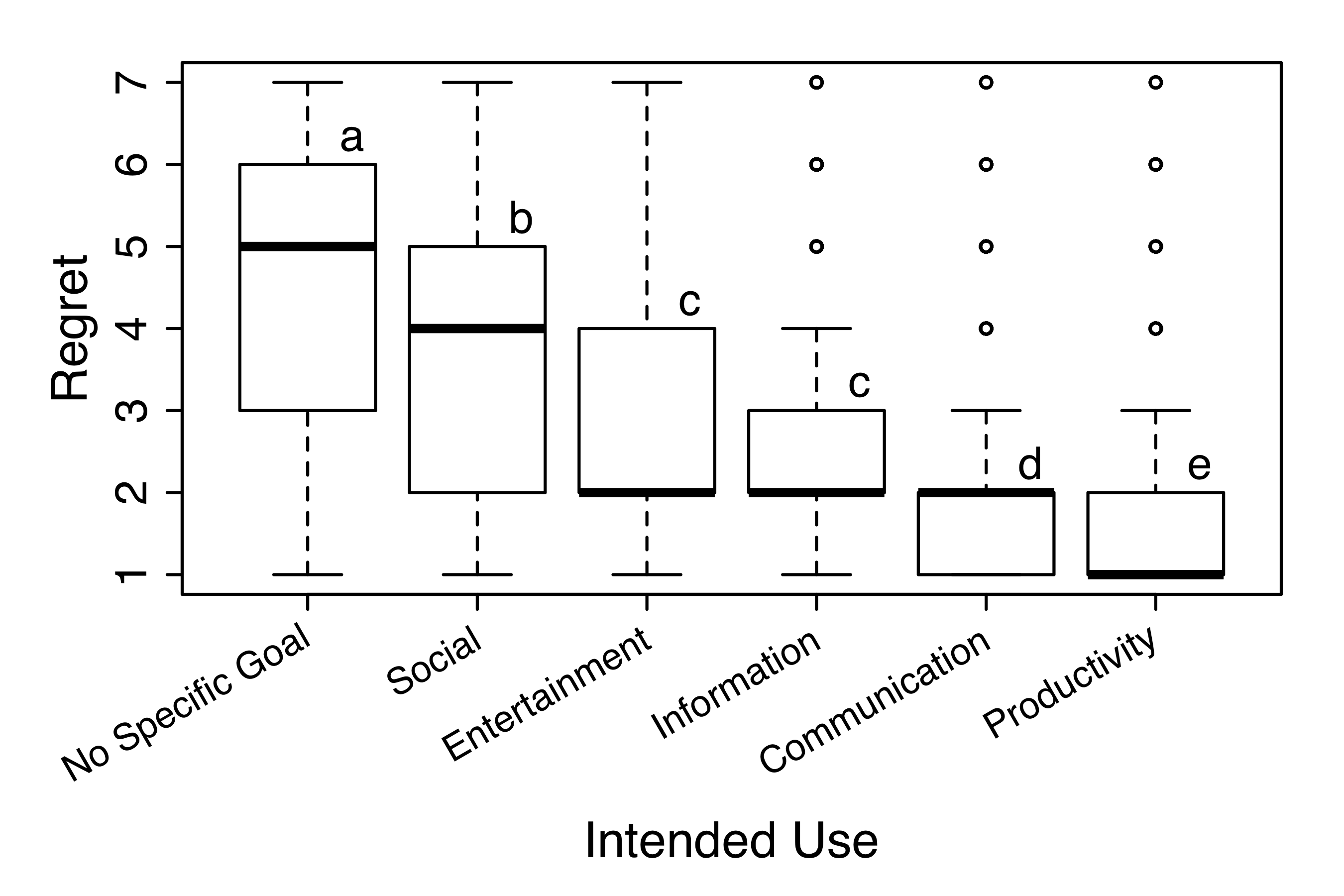}
    \caption{Box plots of \textit{Regret} by \textit{Intended Use}. \textit{Regret} values are responses to the prompt: ``I feel regret about this phone use session'' (1=Strongly Disagree, 7=Strongly Agree). Letters (a, b, c, ...) indicate pairs that are NOT significantly different from each other. For example, we did not find a statistically significant difference between \textit{Entertainment} and \textit{Information}, but did find a statistically significant difference between \textit{No Specific Goal} and \textit{Social}. For pairs with statistically significant difference, $p<.001$ for all of the pairs except between \textit{No Specific Goal} and \textit{Social}, where $p<.01$.}
    \label{fig:regret-intention}
    \Description{A box plot depicting the relationship between “Intended Use” and “Regret”. The x-axis represents different intended uses, including “No Specific Goal”, “Social”, “Entertainment”, “Information”, “Communication”, and “Productivity”. The y-axis represents “Regret” on a scale from 1 to 7. The plot shows that regret is highest for “No Specific Goal” and “Social” use cases and lowest for “Productivity”. Other results are included in the text.}
\end{figure}

We found that the user's stated intention for opening an app predicted the extent to which they would later regret doing so (see Figure~\ref{fig:regret-intention}). An analysis of variance based on mixed ordinal logistic regression indicated a statistically significant effect of \textit{Intended Use} on \textit{Regret} ($p<.001$). Pairwise comparisons (also shown in Figure~\ref{fig:regret-intention}) indicated that participants were significantly more likely to regret their use when it was motivated by \textit{No Specific Goal} than when it was motivated by any other intention (i.e., any of \textit{Communication} ($p<.001$), \textit{Entertainment} ($p<.001$), \textit{Information} ($p<.001$), \textit{Productivity} ($p<.001$), and \textit{Social} ($p<.01$). Post hoc comparisons also revealed that the \textit{Regret} scores for each pair of \textit{Intended Use} except between \textit{Entertainment} and \textit{Information} were significantly different ($p<.001$). 

Overall, these results indicate that participants were most likely to regret phone use when they went to an app without a specific goal, followed by when they wanted to browse social media. Interestingly, participants felt similarly regretful when they were intentionally seeking either entertainment or information, with each being less regrettable than browsing social media. Participants tended to express less regret when their intention in using an app was to communicate or interact with others, and least regretful when they wanted to get things done or self-improve.

\subsubsection{Non-intentional social media use is more regretful than intentional social media use} 
\begin{figure}
    \centering
    \includegraphics[width=\linewidth]{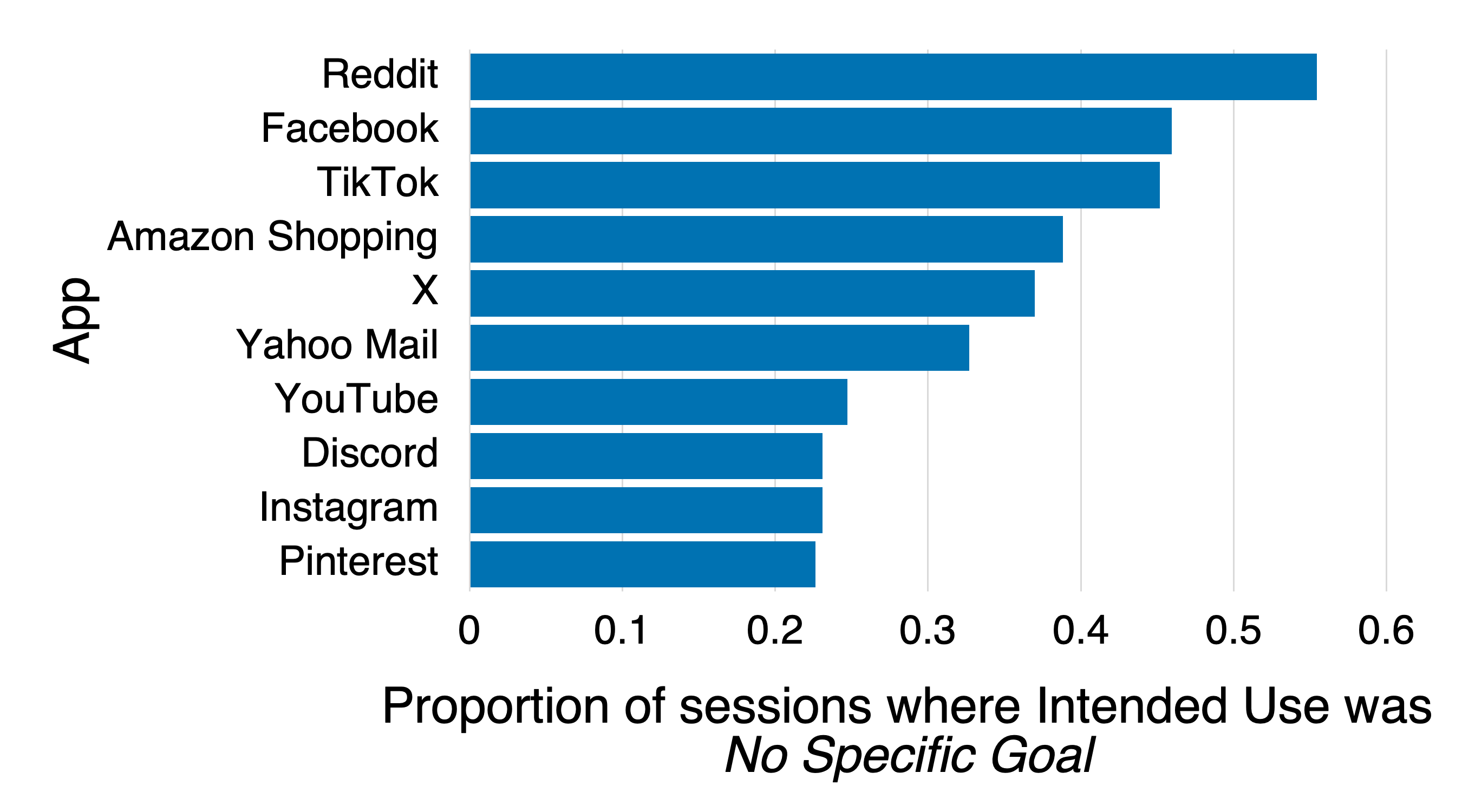}
    \caption{Top 10 Apps with the highest proportion of sessions where user-reported intended use was \textit{No Specific Goal}, after filtering out apps that have been used less than 20 times overall.}
    \label{fig:proportion-none}
    \Description{
    A horizontal bar chart displaying the proportion of sessions where the intended use was categorized as “No Specific Goal” for various apps. The x-axis represents the proportion of sessions, while the y-axis lists the apps, from highest to lowest, including Reddit (between 0.5 and 0.6), Facebook, TikTok, Amazon Shopping, X (formerly Twitter), Yahoo Mail, YouTube, Discord, Instagram, and Pinterest (between 0.2 and 0.3).
    }
\end{figure}

Given that participants were most likely to regret their phone use after using an app without having a goal, we examined which apps they used when they reported not having a goal. Figure~\ref{fig:proportion-none} shows the top 10 apps with the highest proportion of user-reported intention being \textit{No Specific Goal}, after removing apps that have been used less than 20 times overall (due to their overall low frequency in the dataset). Eight of these apps have some form of social media, which aligns with participants' perception, as expressed in interviews, that they usually go to social media apps when habitually checking their phone. For example, P8 said: \textit{``[Instagram] is one of the apps that I just open as a habit for no reason. And I don't even know why.''} (P8, F2)

\begin{figure}
    \centering
    \includegraphics[width=\linewidth]{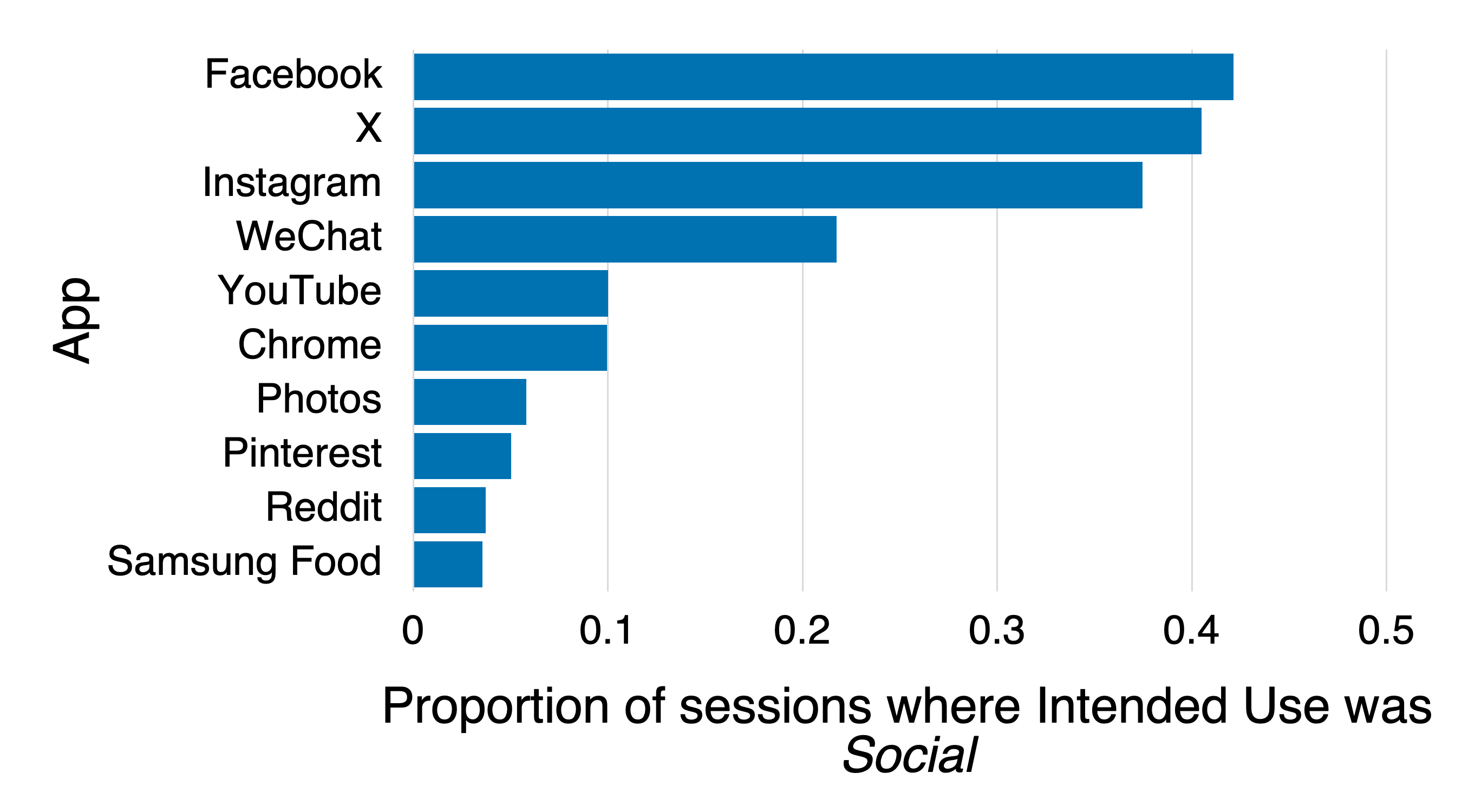}
    \caption{Top 10 Apps with the highest proportion of sessions where user-reported intended use was \textit{Social}, after filtering out apps that have been used less than 20 times overall.}
    \label{fig:proportion-social}
    \Description{
    A horizontal bar chart displaying the proportion of sessions where the intended use was categorized as “Social” for various apps. The x-axis represents the proportion of sessions, while the y-axis lists the apps, from highest to lowest, including Facebook (slightly over 0.4), X, Instagram, WeChat, YouTube, Chrome, Photos, Pinterest, Reddit, and Samsung Food (between 0 and 0.1).
    }
\end{figure}

Although participants were more likely to regret sessions motivated by no specific goal than they were to regret sessions where they intended to browse social media (as described in~\ref{regret-intention}), in both cases they predominantly used social media apps (see Figure~\ref{fig:proportion-none} and Figure~\ref{fig:proportion-social}). Participants explained that they felt good about the times when they intentionally chose to browse social media for pleasure, but they regretted instances where they mindlessly turned to it out of habit or as a knee-jerk avoidance mechanism because of procrastination: 

\begin{quote}
\textit{``I'm gonna go on Facebook and check out social media, but [if] it's like, I'm giving myself a nice break, then I'm less likely to regret it than if I'm just doing it to zone out or to procrastinate an off computer test or an off phone task\ldots [If] I'm doing it because I want to go on social media, then I seem less likely to regret it, than it's like if I didn't really have a specific purpose. I just want to go on my phone to, like, avoid doing other things.''} (P7, F1)
\end{quote}

Thus, both our qualitative and quantitative results suggest that non-intentional, habitual social media use was more regretful than intentional, deliberate social media use.

\subsection{How Regret Varies by Type of Social Media Activity}
\revise{To answer RQ2, w}e next examined participants' fine-grained interactions within social media apps, given that these were the apps they were most likely to regret using. As described in Section~\ref{mtd-screenshot}, we used a multimodal LLM to analyze all screenshots from five social media apps: Instagram, Facebook, Snapchat, TikTok, and Reddit (adding up to 664 app sessions and 34,313 screenshots), evaluating each screenshot for the ``activity'' it represents (one of seven possible choices). 

\subsubsection{Social media activity and regret}\label{activity-regret} 
\begin{figure}
    \centering
    \includegraphics[width=\linewidth]{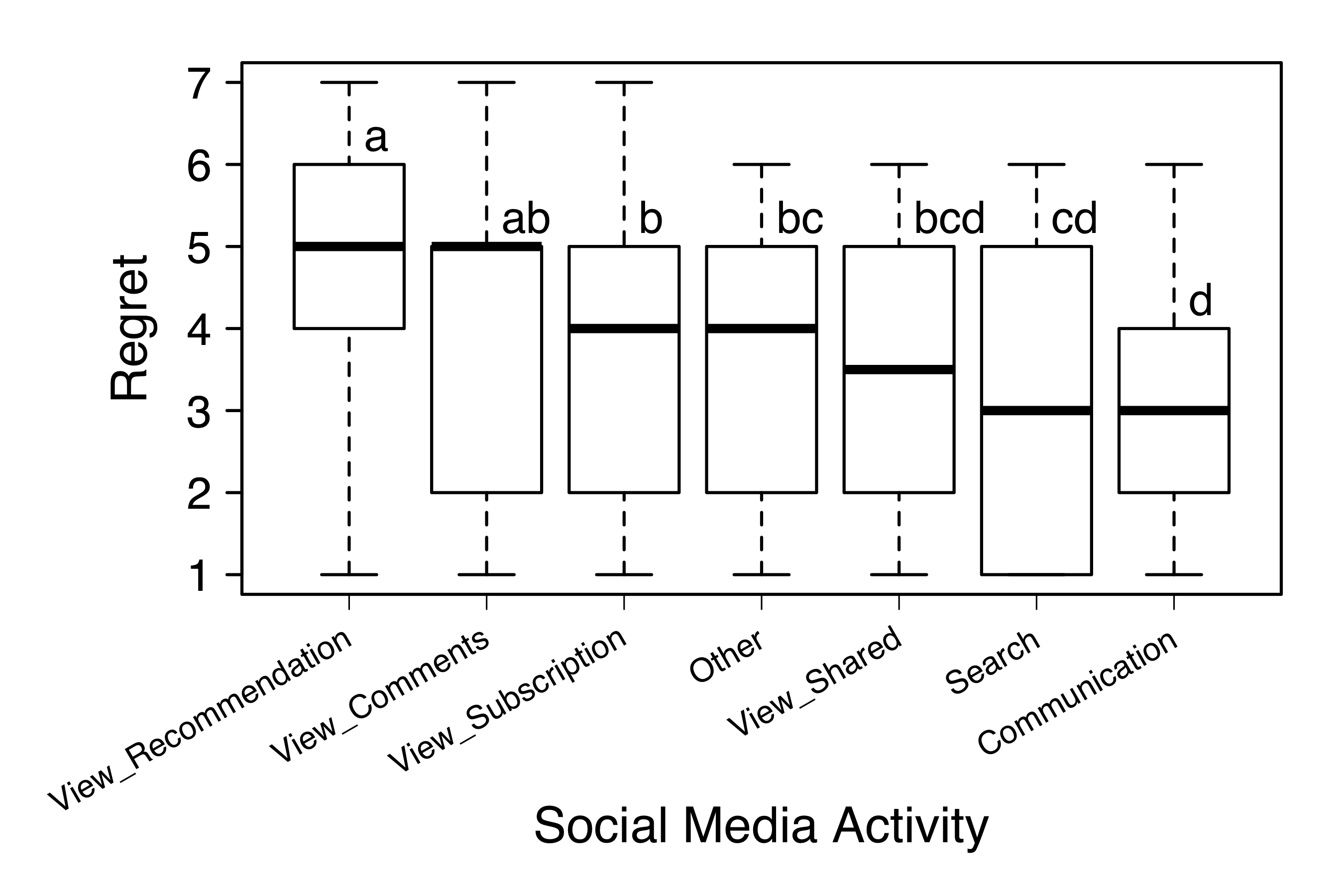}
    \caption{Box plots of \textit{Regret} by \textit{Social Media Activity}. \textit{Regret} values are responses to the prompt: ``I feel regret about this phone use session'' (1=Strongly Disagree, 7=Strongly Agree). Letters (a, b, c, ...) indicate pairs that are NOT significantly different from each other. For example, we did not find a statistically significant difference between \textit{View\_Recommendation} and \textit{View\_Comments}, but did find a statistically significant difference between \textit{View\_Recommendation} and \textit{View\_Subscription}. The ``Other'' category may include anything that does not easily fall into any of the other categories, such as advertisements, notifications, settings menus, etc. \revise{Note that the LLM performance on classifying \textit{View\_Shared} was low (often confused with \textit{View\_Subscription}), so the regret associated with it may not be accurate.}}
    \label{fig:activity-regret}
    \Description{
    A box plot illustrating the relationship between “Social Media Activity” and “Regret”. The x-axis represents different types of social media activities, including “View_Recommendation”, “View_Comments”, “View_Subscription”, “Other”, “View_Shared”, “Search”, and “Communication”. The y-axis represents “Regret” on a scale from 1 to 7. Each box plot displays the median, interquartile range, whiskers, and outliers. Letters above the boxes indicate statistically significant differences, with “a” representing the highest regret and “d” the lowest. Viewing Recommendation-Based Content (e.g., algorithmically suggested posts) is associated with the highest regret, significantly more than most other activities except View_Comments. Communication activities have the lowest regret, showing a significant difference from other activities except search. Other results are included in the text.
    }
\end{figure}

To understand how regret is related to different social media activities, we tested the relationship between these two variables. Using the most prevalent activity in each app session to represent the session, we conducted an analysis of variance based on mixed ordinal logistic regression, which indicated a statistically significant effect of \textit{Activity} on \textit{Regret} $p<.001$. Pairwise comparisons indicated that regret Likert scores for \textit{View\_Recommendation} vs. all the other activities except \textit{View\_Comments}, including \textit{View\_Subscription} ($p<.01$), \textit{Other} ($p<.001$), \textit{Search} ($p<.001$), and \textit{Communication} ($p<.001$), were statistically significantly different, that the regret scores for \textit{View\_Comments} vs. \textit{Communication} ($p<.001$) and \textit{Search} ($p<.01$) were statistically significantly different, that the regret scores for \textit{View\_Subscription} vs. \textit{Communication} ($p<.001$) and \textit{Search} ($p<.05$) were statistically significantly different, and that the regret scores for \textit{Other} vs. \textit{Communication} ($p<.01$) were statistically significantly different. Figure~\ref{fig:activity-regret} shows the box plots of regret scores for each activity category, along with results of post-hoc pairwise comparisons. These results indicate that viewing recommendation-based content and comments or discussion threads are the most regrettable experiences of social media for the participants, followed by viewing subscription-based content, while communication, and search are less regrettable.

These results align with what the participants shared in the follow-up interviews. Many participants said that not all social media usage was equally regrettable to them. Specifically, participants said that activities like direct messaging were meaningful and not regrettable at all because of the human connection they brought. For example, P7 said:

\begin{quote}
    \textit{``One thing I noticed during this research study is I almost never regretted, like Messenger or phone call or like something where I'm interacting with another person that I know\ldots It feels meaningful.''} (P7, F2)
\end{quote}

When it comes to passively consuming social media content,  participants said that the source of the content tends to lead to different levels of regret. Specifically, some participants said that recommendation-based content was more regrettable than subscription-based content. For example, one participant described recommendation-based content as ``random'' and more regrettable than subscription-based content and content they intentionally searched for:

\begin{quote}
    \textit{``So random content, like what Twitter gives me or Facebook. Like almost all of it was random. It wasn't people I knew, right. So like that content, I end up regretting more than if it's content of like people I know or a work thing, like an article I was reading. [\ldots] Even content that I specifically look for, I regret a little less I think than the random stuff.''} (P11, F1)
\end{quote}

Another participant shared a similar perspective, adding that they had short attention span when looking at recommended videos:

\begin{quote}
    \textit{``So it's very random content. Like it's always changing a lot\ldots It's not only changing videos, it's changing content like styles or genres\ldots My attention span is short and I'm not interacting with the videos and the content is very random. So I'm like, willingly engage, like staying in this place where the content isn't like valuable.''} (P9, F2)
\end{quote}

Some participants mentioned that reading comments on social media, especially political ones, can be a negative experience to them: \textit{``When I'm in the news, sometimes when I go into comment sections, then I'd be more likely to regret the content because especially with political topics, there's so many obnoxious comments in the comment section.''} (P7, F2)

Thus, both of our quantitative and qualitative results show that there are systematic patterns linking specific social media activity to regret. Behaviors like consuming content that is algorithmically recommended are associated to higher regret and are perceived as less valuable and relevant by our participants, whereas they found active communication to be meaningful. 

\subsubsection{Visualizing social media activity over the duration of an app session} 
\begin{figure*}
    \centering
    \includegraphics[width=\linewidth]{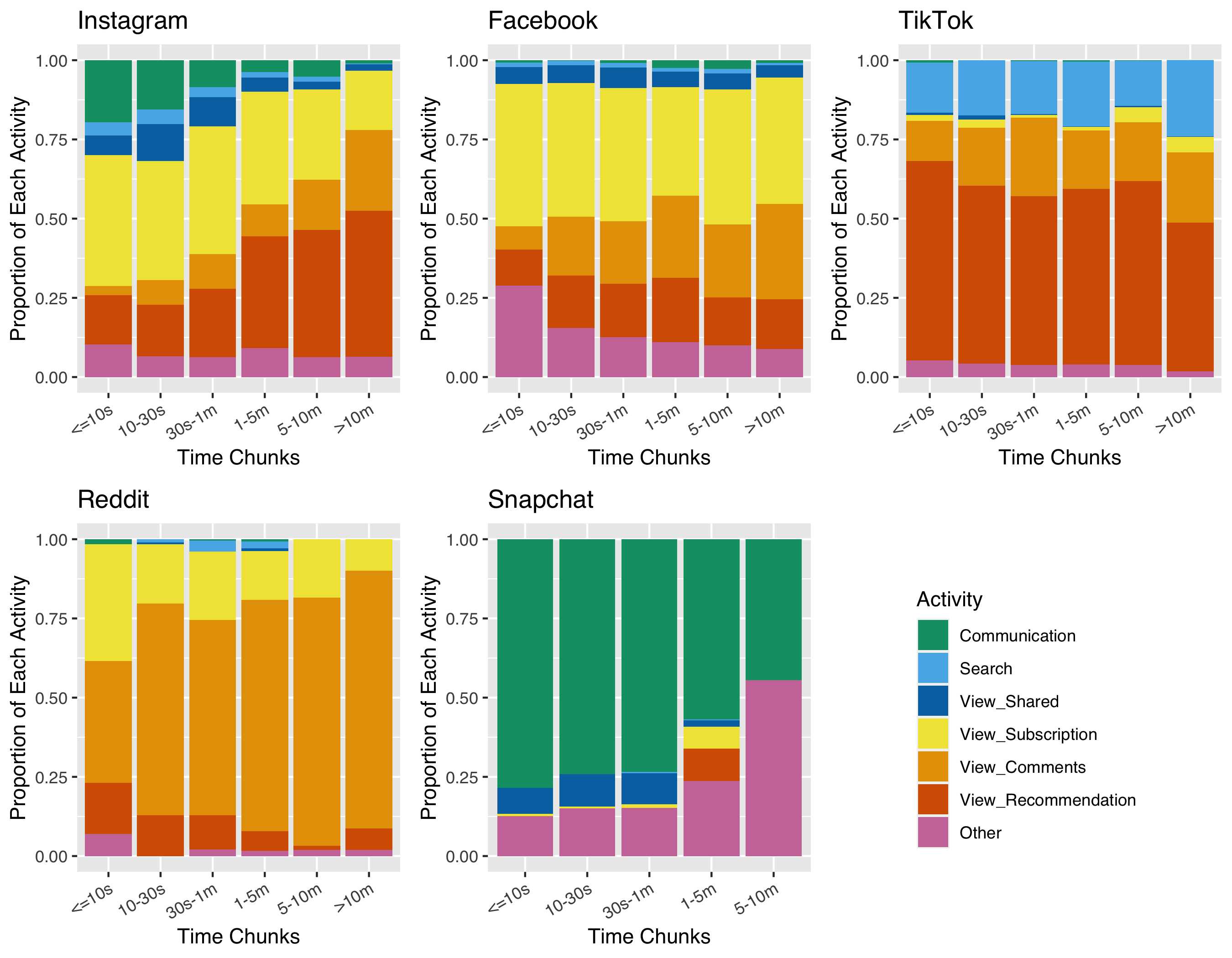}
    \caption{Activity in each social media app in time chunks. Each time chunk contains proportions of each activity in that specific time frame. \revise{Note that the LLM performance on classifying \textit{View\_Shared} was low (often confused with \textit{View\_Subscription}), so its representation in the timeline it may not be accurate.}}
    \label{fig:activity-proportion}
    \Description{
    A set of five stacked bar charts displaying the proportion of different social media activities over various time chunks for Instagram, Facebook, TikTok, Reddit, and Snapchat. The x-axis represents time chunks (ranging from ≤10 seconds to >10 minutes), while the y-axis represents the proportion of each activity. Each bar is color-coded according to different activity categories. TikTok shows a dominant proportion of View_Recommendation activity across all time chunks. Facebook and Instagram display a mix of View_Recommendation, View_Subscription, and View_Comments. On Instagram, View_Recommendation and View_Comments increase in later time chunks, while Communication and View_Subscription decrease. Snapchat differs from other platforms, with Communication being the primary activity across all time chunks. Search and View_Shared activities are relatively low across platforms, except for small proportions in Facebook and Instagram. Other activities are present in all platforms but are relatively low in proportion.
    }
\end{figure*}

Given that people's regret varies based on specific activities on social media, we sought to understand how those activities are distributed in different apps and over the duration of an app session. We adopted the approach of segmenting sessions based on time elapsed within every app session in~\cite{cho2021} and created a timeline for each of the five social media apps we examined, with each showing the breakdown of users' activity at each time chunk of the app session (see Figure~\ref{fig:activity-proportion}). Each timeline revealed a distinctive behavioral signature in the way users engaged with the app. For example, on Instagram, participants tended to begin their usage session by reading and sending direct messages and viewing content from accounts they follow. After approximately one minute, the balance of activities began to shift, and participants increasingly viewed recommended content from accounts they do not follow. In contrast, on Reddit, participants tended to begin by browsing their feed (which consisted of a mix content from accounts they follow and other recommended content from accounts they do not follow) at first, and then dove into specific discussion threads. On TikTok, participants mostly consumed recommended content from accounts they do not follow throughout the duration of the session.

These timelines show that different apps consist of different percentage of each type of social media content, and there are distinct patterns of how users move between regretful and less regretful activities, suggesting that the design of these social media apps can influence user's tendency to engage in behaviors they would later regret. For example, the start of an Instagram session tends to be filled with low regret activities, such as direct messaging and viewing content from followed accounts, but users slowly shift into less valuable activities as their sessions went on. Reddit users spend the majority of their time reading other users' comments, a choice we found to be highly regrettable. TikTok users consistently engaged with a high proportion of recommended content, the most regrettable behavior among our seven categories, from the moment they were on the app.

\subsubsection{Factors that predict regret} To understand what factors can predict regret, we fit a cumulative logistic regression model. We modeled the ratio of each activity within a session (from 0 to 1) and \textit{Intended Use} as fixed effects and \textit{User ID} as a random effect. We added \textit{Duration} and \textit{App} as fixed effects to control for the effect of the time spent on each app session, as most participants reported that it is an important factor that influences their regret. 



We found that \textit{Duration} ($p<.001$), \textit{Intended Use} ($p<.001$), \textit{Ratio\_View\_Recommendation} ($p<.001$), \textit{Ratio\_View\_Comments} ($p<.01$), and \textit{Ratio\_View\_Subscription} ($p<.05$) all had a significant effect on \textit{Regret}. These results show that the amount of time spent on an app, participant's intention when entering the app, and the amount of time they viewed recommendation-based content, comments, and subscription-based content can predict their subsequent regret. Notably, we did not find a significant effect of \textit{App} on \textit{Regret} ($p=.058$), indicating that after controlling for intended use, time spent on app, and specific app activity, the app itself does not predict regret. These results align with our results in \ref{regret-intention} and \ref{activity-regret}, and they further show that user intention and the amount of recommendation-based, subscription-based content and comment consumption are all individually important predictors of regret when holding other factors constant.

\subsection{How Regret Is Influenced by Deviation from Communication to Browsing Social Media}

Given that the two variables we examined above, intended use and social media activity, indicate what people want to do and what people actually do respectively, a natural hypothesis \revise{(connected to RQ3)} that arises is that people may feel more regret when their actual behavior deviates from their original intention than when the behavior aligns with their intention, as suggested by prior work~\cite{cho2021} \revise{and the theory on regret regulation, which states that intention-behavior inconsistency increases process regret~\cite{zeelenberg2007, pieters2005}}. Indeed, participants mentioned that when they deviated from their initial purpose and stayed longer on the app, it made them feel regret: \textit{``That's when I put somewhat agree to regretting\ldots because it starts off with good intentions with like, I'm just trying to entertain myself, and then it ends with like, okay, let's hear a little bit longer than I should have.''} (P2, F1)

However, what specific activity counts as ``deviation'' from an initial purpose can be challenging to pinpoint. For example, if participants reported wanting to get information, they may or may not want to get information from accounts they follow or through active search. To investigate this pattern of deviation, we chose a specific case where the mapping between intended use and activity is clear: when participants reported wanting to passively browse social media or actively communicate or interact with others. In the follow-up interviews, participants brought up that when they started with an initial purpose to communicate, sometimes they ended up browsing on their feed instead, a choice they later regretted. For example, when reviewing a session in a follow-up interview, P3 said: \textit{``I think I planned to communicate with someone but I didn't. I lost my purpose.''} (P3, F1) Similarly, P4, at the end of the last follow-up interview, summarized this as an indicator of regret: \textit{``Even if I spend a session on Instagram and I do DM at the beginning of that session, like that's a good indicator that I'm not gonna regret it. But then if I only do that for like 30 seconds and then I spend half an hour on the app, then I'm much more likely to regret it.''} (P4, F2)

\begin{figure}
    \centering
    \includegraphics[width=\linewidth]{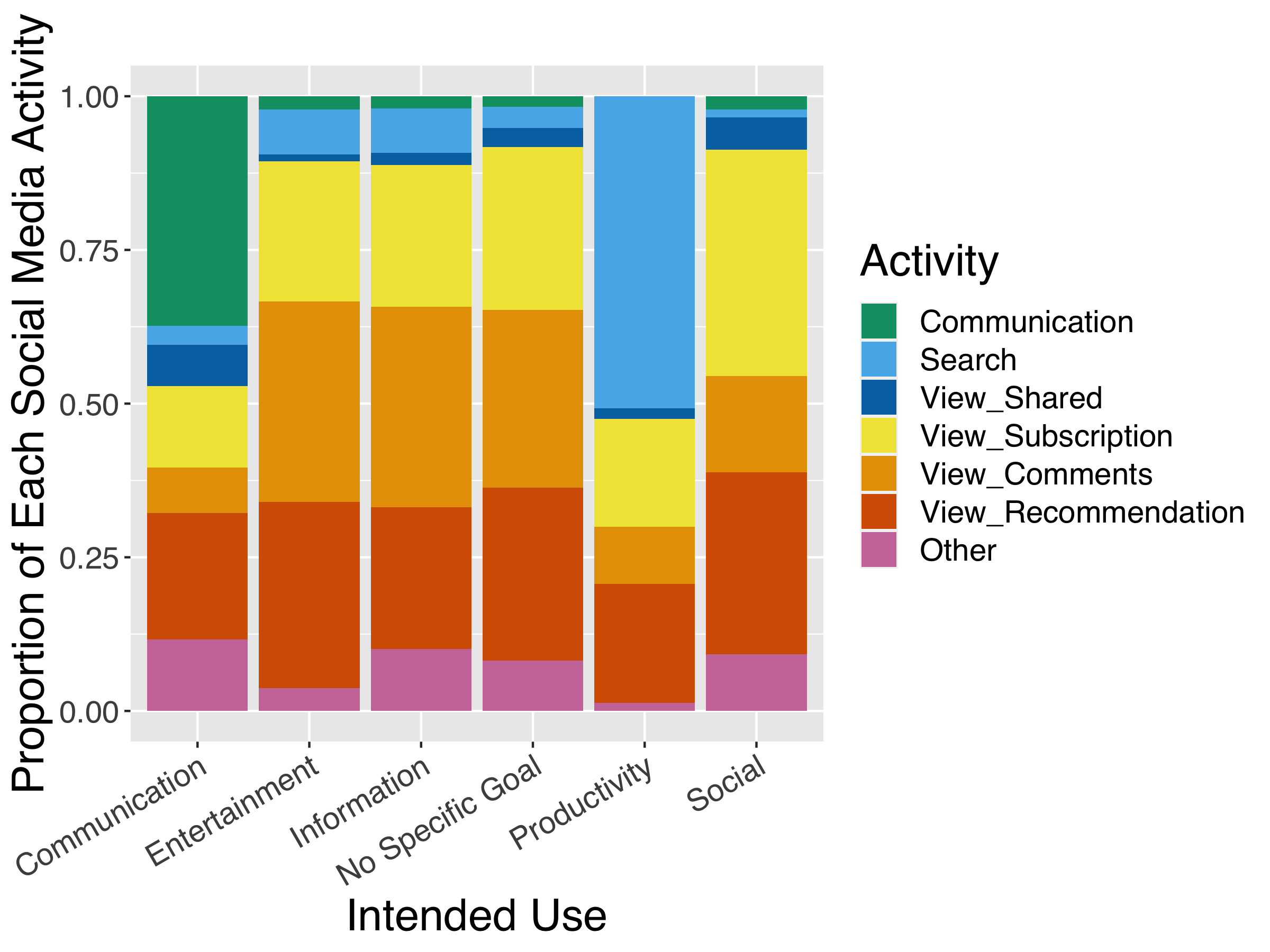}
    \caption{Proportion of social media activity for each intended use \revise{(which participants provided \textit{prior to} entering every social media app). Note that the LLM performance on classifying \textit{View\_Shared} was low (often confused with \textit{View\_Subscription}), so the proportion associated with it may not be accurate.}}
    \label{fig:proportion-activity-intention}
    \Description{
    A stacked bar chart illustrating the proportion of different social media activities across various intended uses. The x-axis represents intended uses, including “Communication”, “Entertainment”, “Information”, “No Specific Goal”, “Productivity”, and “Social”. The y-axis represents the proportion of each social media activity within each intended use category. The activities are color-coded. Communication activity is dominant in the Communication intended use, but only accounts for 30 to 40\%. Search activity is prominent in Productivity, distinguishing it from other intended uses. Entertainment, and Social. Entertainment, Information, No Specific Goal, and Social have similar distributions, with high proportions of View_Recommendation, View_Comments, and View_Subscription activities. Other activity remains relatively low across all categories.
    }
\end{figure}

To quantify this deviation, we first visualized the proportion of screenshots associated with each activity within each intended use (see Figure~\ref{fig:proportion-activity-intention}). In sessions where users started with an intention to communicate or interact with others (the first bar in Figure~\ref{fig:proportion-activity-intention}), 37.3\% of the screenshots were labeled as \textit{Communication} by LLM, the rest included \textit{View\_Recommendation} (20.5\%), \textit{View\_Subscription} (13.3\%), \textit{Other} (11.6\%), \textit{View\_Comments} (7.4\%), \textit{View\_Shared} (6.7\%, \revise{although this might be inaccurate due to mislabeling}), and \textit{Search} (3.1\%). In sessions where users started with an intention to browse social media (the last bar in Figure~\ref{fig:proportion-activity-intention}), 36.8\% were \textit{View\_Subscription}, 29.6\% were \textit{View\_Recommendation}, 15.7\% were \textit{View\_Comments}, 9.2\% were \textit{Other}, 5.2\% were \textit{View\_Shared}, 2.2\% were \textit{Communication}, and 1.3\% were \textit{Search}. These results suggest that while participants frequently deviated to non-communication behavior when having an intention to communicate (over 60\% of the time), they rarely started with an intention to browse social media but ended up communicating (around 2\% of the time).

\begin{figure}
    \centering
    \includegraphics[width=\linewidth]{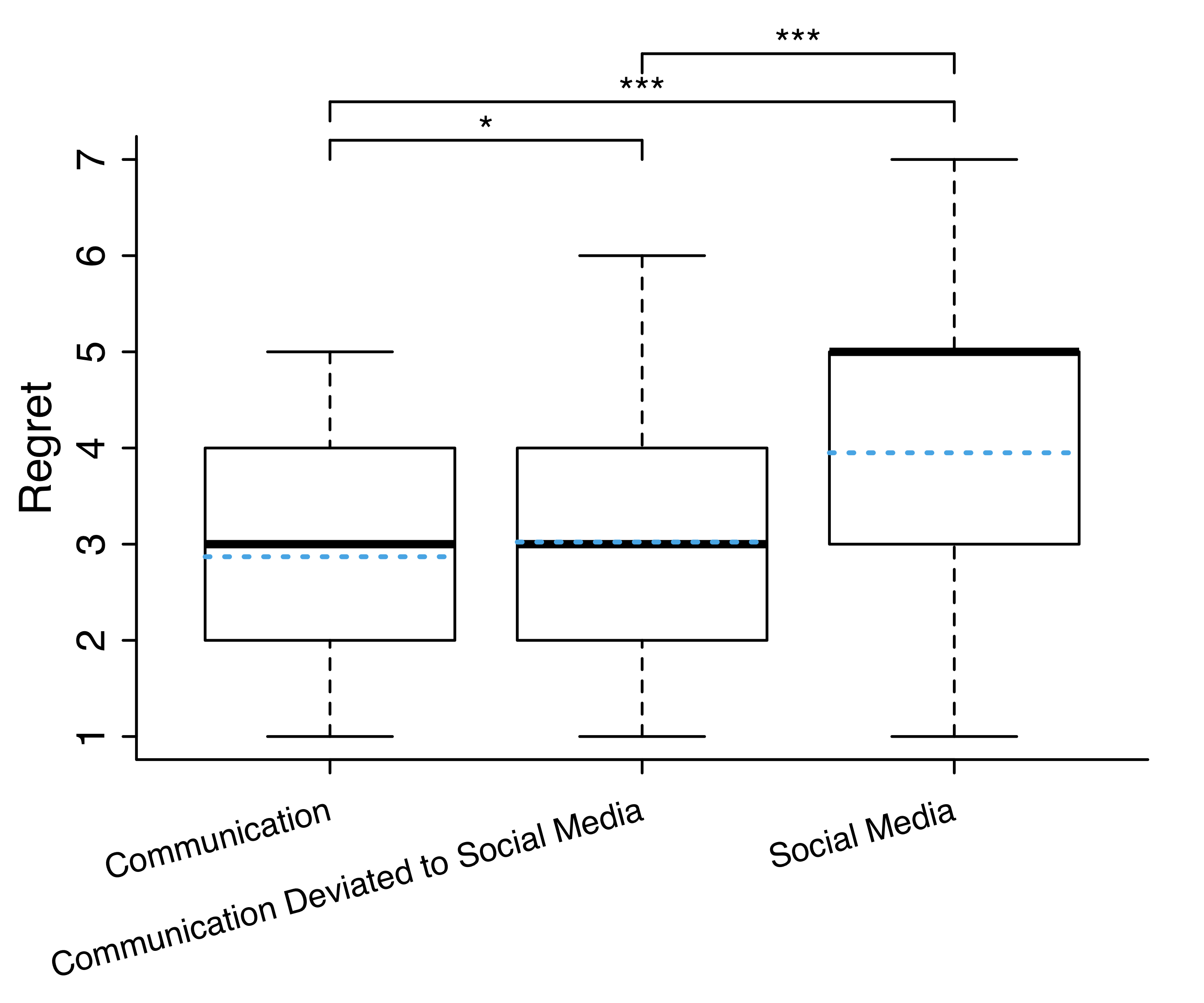}
    \caption{Box plots of Regret for three groups. The first group, labeled as ``Communication,'' includes sessions where the user intention was to communicate or interact with others, and the subsequent activity (from the screenshots) shows communication only. The second group, labeled as ``Communication Deviated to Social Media,'' includes sessions where the user intention was to communicate or interact with others, but the subsequent activity includes things other than communication, suggesting the user has deviated from their intention to browsing social media. The last group, labeled as ``Social Media,'' includes sessions where the user intention was to browse social media. Significant levels of difference between each pair are indicated in the plot, where * means $p<.05$ and *** means $p<.001$. The blue dashed line in each box plot represents the mean for that group, and the black solid line represents the median.}
    \label{fig:deviation}
    \Description{
    A box plot comparing Regret across three categories: “Communication”, “Communication Deviated to Social Media”, and “Social Media”. The x-axis represents the different types of activities, while the y-axis represents regret on a scale from 1 to 7. The solid black line inside each box represents the median, while the blue dashed line represents the mean. Communication and Communication Deviated to Social Media have similar distributions, with the second group having a slightly higher mean value and a max of 6 (vs. 5). Social Media has higher mean, median, and upper and lower quartile than the other two groups. Other results are reported in the text.
    }
\end{figure}

To examine how deviation from communication to browsing social media may be related to regret, we set up our data analysis in the following way. We extracted three groups of sessions from our dataset. The first group includes sessions where the reported intended use was \textit{Communication}, and the only activity from that session was \textit{Communication} (i.e., LLM identified direct communication from all the screenshots). The second group includes sessions where the reported intended use was \textit{Communication}, but activities other than direct communication were present (i.e., the user deviated from there intention to communicate). As a comparison, we also added a third group, which includes sessions where the intention was to browse social media. Figure~\ref{fig:deviation} shows box plots of \textit{Regret} for these three groups. These three groups add up to 317 app sessions, with each group having 46, 90, 181 sessions respectively. We found a statistically significant difference of \textit{Regret} between the three groups ($p<.001$). Pairwise comparisons indicated that the \textit{Communication} and \textit{Communication Deviated to Social Media} groups were statistically significantly different ($p<.05$), and that both groups were statistically significantly different from \textit{Social Media} ($p<.001$). The median regret scores for \textit{Communication} and \textit{Communication Deviated to Social Media} are both 3, whereas the median regret score for \textit{Social Media} is 5. As another measure of central tendency, average regret scores for each group are 2.87, 3.02, and 3.95.

These results indicate that participants tended to feel just slightly more regretful when they deviated from active communication to passive browsing of social media. But regardless of whether they deviated, sessions where the user intention was to browse social media was much more regretful than sessions where the intention was to communicate or interact with others.

\section{Discussion}

\subsection{Characterizing Regrettable \revise{Social Media} Use}\label{chracterize-regret}
Through our mixed-methods analysis, our results robustly showed that people's intentions when using an app is an important predictor of later regret. We consistently saw that non-intentional phone use is related to an increased sense of regret later on. This aligns with prior work suggesting that ritualistic use (i.e., where the user habitually uses their phone to pass the time) is associated with a lower sense of meaning~\cite{lukoff2018}. This effect persists even after holding other factors constant: what app the user is using, the specific activity they engage in, and how much time they spend on the app. \revise{Theories of regret explain that people regret actions that are less justifiable and those that stem from poor decision-making processes~\cite{zeelenberg2007, gilovich1995}. This may explain why people experience more regret over phone-use sessions that are not motivated by specific user intentions. Picking up the phone for no specific reason is harder to justify than doing so in pursuit of a specific intention.} Cho et al. argue that people regret their phone use when it reflects the impulsivity of choosing a small, short-term reward (e.g., browsing their phone) over a bigger one (e.g., engaging in productive activities)~\cite{cho2021}. This can also explain why people regret moments of non-intentional phone use, as this non-intentional use is often characterized by a lack of purpose and habitual or reflexive behavior, in which the user may be prone to impulsiveness. Given the importance of the user's intention in shaping their regret, what remains to be understood is the precise nature of such non-intentional use. For example, participants in our study said that they felt the urge to engage in non-intentional use when they were procrastinating and trying to avoid doing other activities. How other factors (such as the user's own mental state, their context, and design factors like push notifications) might trigger such non-intentional use is worth exploring in future work.

Our findings also highlight the benefits of examining the different ways that people engage with social media, rather than treating all social media use holistically. We found that people's regret can vary depending on the dominant activity in a social media usage session; participants were most likely to feel regret after viewing content that was algorithmically recommended to them or after browsing comments or discussion threads. And they felt least regretful after they had communicated with other people. These findings contribute to the ongoing debate on how passive and active social media use contributes to well-being~\cite{active-passive-media}, providing evidence that active communication (through direct messaging) is associated with lower negative affect than other activities.

\revise{Our findings can be interpreted both in terms of outcome regret and process regret. In terms of outcomes, Cho and colleagues explain that users seek four types of rewards from social media use, including: social rewards, informational rewards, personal interests, and entertainment rewards~\cite{cho2021}. Activities like searching for information and direct messaging may offer particularly high informational rewards and social rewards respectively, and participants described these activities as more ``meaningful.'' This data suggests they tend to induce less outcome regret than other activities.} 

\revise{In considering process regret, it is worth noting that when recommended content is displayed to users by a social media algorithm (without users explicitly choosing to view this content), this process can feel less justifiable, which induces more process regret. Indeed, in the interviews, participants described such content as ``random,'' indicating that they often run into recommended content in an arbitrary, unplanned way, without a conscious or justified reason stemming from a well-considered decision-making process. These two independent sources of regret may jointly contribute to the higher regret associated with viewing algorithmically recommended content and comments, and the lower regret associated with active search and communication.} 

Past work has found through interviews that when users ``sidetrack,'' deviating from their original intentions, they experience more regret about their usage~\cite{cho2021}. \revise{This is supported by Zeelenberg and Pieters' work on regret regulation, which shows that intention-behavior inconsistency increases regret~\cite{zeelenberg2007, pieters2005}.} In this study, we examined this phenomenon within the context of a specific usage case, where users' behavior deviates from their intention to communicate or interact with people. We characterized the prevalence of such deviation, finding that when participants intended to communicate with other people, 60\% of the time they deviated from this intention and browsed social media. Although this deviation was significantly associated with increased regret relative to non-deviation sessions, this increase was slight. And these sessions were far less regretful than sessions where users set out to browse social media in the first place. One potential explanation for this result is that if people start with an intention they judge positively (such as communicating with a friend), and they go on to fulfill this intention, they may feel satisfied with their experience overall, even if they later deviate to passive browsing. As noted earlier, we chose a specific form of deviation in this study, and future work can explore alternative forms to further model the relationship between intention-activity deviation and regret (e.g., by analyzing how activity changes and deviates throughout an app usage session).

\subsection{Implications for Design and Policy}
We encountered systematic patterns linking particular behaviors with later feelings of regret (see Figure~\ref{fig:activity-regret}) and linking the design of an app with particular behaviors (see Figure~\ref{fig:activity-proportion}). For example, Instagram users began their sessions by viewing content from people they follow but were gradually funneled into viewing algorithmically recommended content from unknown accounts; Reddit users gradually descended into comment black holes; TikTok users engaged with mostly recommended content the moment they landed on the app. All of these patterns reflect a shift away from behaviors users value and toward behaviors they do not.

These findings have important implications for regulating attention-economy designs. Regulation of the technology industry strives to protect users from harm, including deceptive and manipulative practices that encourage users to act against their own best interests~\cite{communicationacm}. Historically, this work has targeted financial and data-privacy harms (e.g., ~\cite{mathur2019, bosch2016tales}), identifying, for example, manipulative designs that encourage users to make purchases they do not need or intend to make. A growing body of academic work has called for the investigation of attentional harm~\cite{switchtube} and shown that attention-capture designs are widespread~\cite{roffarello2023}. However, demonstrating manipulation and harm has been comparatively challenging with respect to users' engagement, because the fact that a user chooses to engage is argued to be a reflection of their preferences~\cite{richards2024against}. Our results provide continuing evidence that users' engagement decisions are not always a reflection of their intrinsic desires, but instead \revise{reflect a lack of justifiability that increases regret and goes} against their best interests.

\begin{figure}
    \centering
    \includegraphics[width=\linewidth]{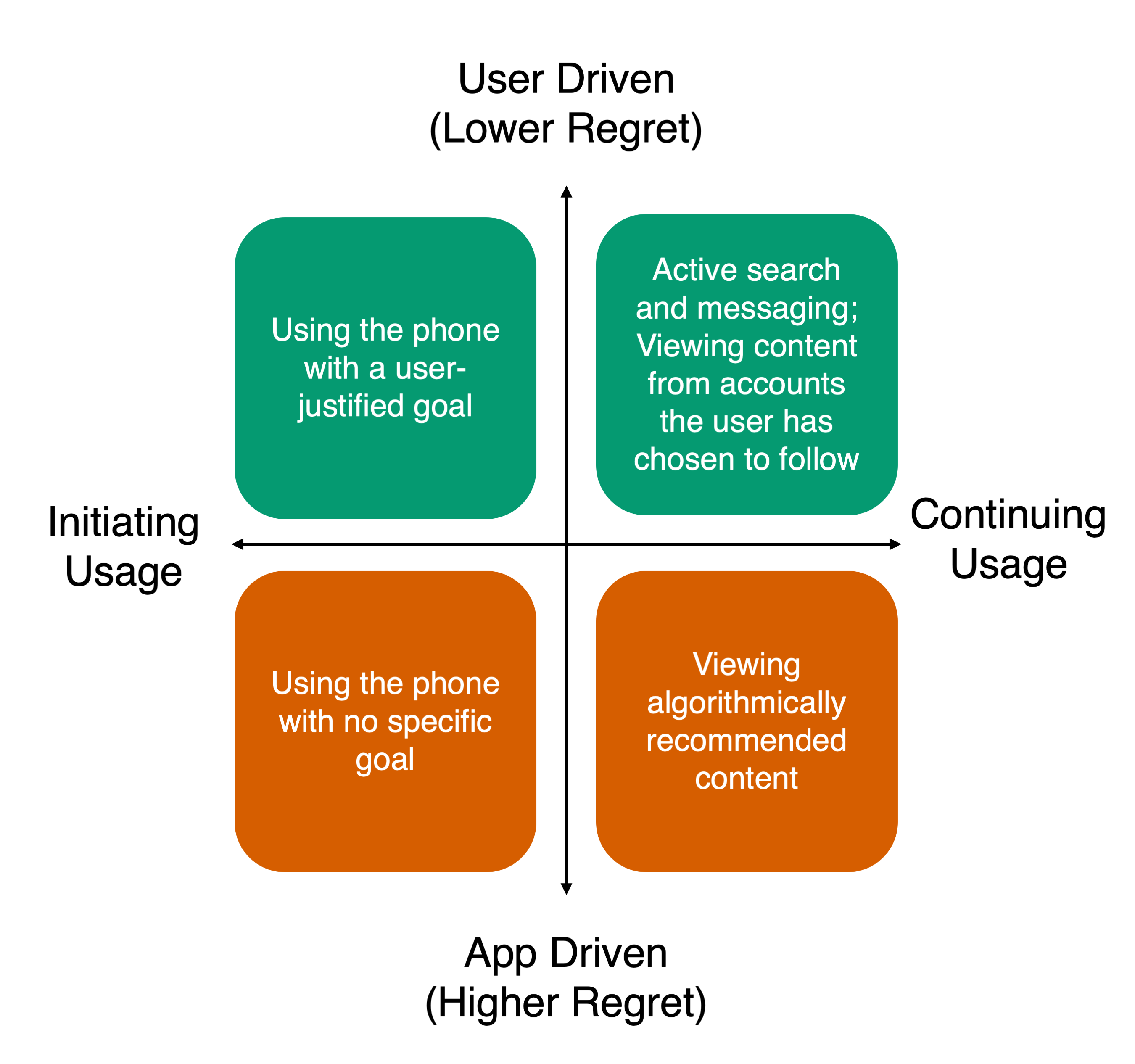}
    \caption{\revise{Regret matrix. We divide regretful behavior along two axes: app or user driven and the stage of usage (initiating or continuing usage). The two green blocks indicate experiences associated with lower regret, while the two orange blocks indicate experiences associated with higher regret.}}
    \label{fig:regretMatrix}
    \Description{
    A quadrant diagram categorizing phone usage behaviors based on user-driven vs. app-driven engagement and initiating vs. continuing usage. The x-axis represents the transition from initiating usage (left) to continuing usage (right), while the y-axis represents the spectrum from user-driven (lower regret, top) to app-driven (higher regret, bottom) behaviors. Top-left (User-Driven, Initiating Usage): “Using the phone with a user-justified goal.” Top-right (User-Driven, Continuing Usage): “Active search and messaging; Viewing content from accounts the user has chosen to follow.” Bottom-left (App-Driven, Initiating Usage): “Using the phone with no specific goal.” Bottom-right (App-Driven, Continuing Usage): “Viewing algorithmically recommended content.”
    }
\end{figure}

Further, our findings point to a clear link between \revise{(process)} regret and engagement decisions that are app-driven (see Figure~\ref{fig:regretMatrix}). When participants' decision to initiate engagement came from an intrinsic goal that they were able to articulate, they were far less likely to regret their use than when they could not come up with a reason for why they were there. And once engaged, they were far less likely to regret staying engaged when they looked at content and messages from people they know and have chosen to follow; they were much more likely to regret staying engaged when the app pushed content of its own choosing. \revise{These findings have important implications for social media apps. To maximize users' well-being and reduce aversive feelings such as regret, app designers should not only consider the outcome of a user's engagement (for example, whether they will value the content being recommended to them), but also aim to reduce process regret that stems from unjustified, unplanned, and unconscious use. For example, apps could explicitly assess the user's intention upon app entry, adjust the interface to respect that intention, and avoid interspersing features and content in a way that may distract users and lead to sidetracking. The SwitchTube system, which introduced the notion of ``adaptive commitment interface,'' is an example of such an experience, where the interface aligns with the user's goal~\cite{switchtube}.} These findings also point to a need for tools and benchmarks that measure a user's intention as they engage with an interface. If users consistently lose the thread of their intention, the app in question could benefit from redesign and regulation. Regulations that require companies to assess users' intentionality and to remove features that undermine it have the potential to reduce attentional harm and designs that funnel users into behaviors they regret. 

\subsection{Using Screenshots and Multimodal Large Language Model to Understand Phone-Use Behavior}
In this study, we introduced a novel method to analyze fine-grained phone-use behavior through passive screenshot collection and automatic analysis using an MLLM. We showed that this method achieved a substantial agreement (a Cohen's Kappa of 0.736) with human consensus labels, an accuracy of 79.5\%, and a weighted average F1 score of 79.7\%. Compared to previous methods that similarly analyzed fine-grained behaviors on social media platforms using the Android Accessibility API (e.g., \cite{cho2021, orzikulova2023}), our approach has a few advantages. First, our approach does not require access to an app's UI structure during data collection and therefore does not require a strict rule-based mechanism to determine the feature that the user is interacting with. Second, our approach is more robust against app updates and can be applied to a wide range of apps more easily by simply changing the prompt to LLM. However, one limitation of our approach compared to using the Accessibility API is that LLM performance can degrade when certain categories share similar visual features with one another. For example, in this study, the LLM classification performance for viewing shared content from friends was much lower than for other categories. In this study, we employed zero-shot chain-of-thought prompting when using an MLLM. It would be valuable for future work to explore other ways to leverage MLLMs to improve the classification performance, potentially through few-shot chain-of-thought prompting~\cite{chain-of-thought} and vision fine-tuning~\cite{openai2024vision}.

For researchers interested in adopting our approach, we generally recommend keeping a short window of screenshots (e.g., around five screenshots) rather than one screenshot during categorization (both for human and LLM), since there might be details missing from just one screenshot. For example, if the user scrolled past the title of a post, it might be hard to tell whether the content is from someone they followed or the platform's recommendation algorithm. Having screenshots from seconds ago can be helpful in such cases. In this study, we used text descriptions of five screenshots (25 seconds) for our categorization task. In rare instances, we still found that we needed more screenshots to be able to classify correctly. For example, if a user opened shared content from a conversation, it might be necessary to trace back to earlier screenshots to see the origin of the content. However, we noticed that adding too many screenshots as context can degrade the current model's performance (e.g., more than 10 screenshots), although we recognize that this might change as models improve.

We note that collecting raw screenshots of phone use may present privacy concerns for participants. It would be valuable for future work to investigate privacy-preserving approaches to extract relevant information from real-time screenshots on-device. Past work has shown that on-device models can be trained to detect UI elements~\cite{zhang2021} and extract information from screenshots~\cite{kumar2020}. Similar approaches could potentially replace the first step of our screenshot analysis process (in Figure~\ref{fig:llm}) and make the pipeline more privacy-preserving.

While our approach was developed for this empirical study, we believe that it can also be adopted for smartphone overuse intervention systems. Given that screenshots contain rich information about a user's activity, future work can explore combining screenshot analysis with passive sensing data to predict problematic phone use and deliver adaptive interventions tailored to users' in-the-moment contexts. For such a system to be deployed in the wild, privacy is an important consideration, and on-device approaches mentioned above might be a promising solution to alleviate privacy concerns. For a just-in-time intervention system, speed is also an important factor to consider. In our study, we found that retaining 1/16 of the original resolution of the screenshots and taking screenshots every 5 seconds was an ideal combination, but a just-in-time system might have a higher threshold for speed and responsiveness. Future work can further explore variations of the screenshot collection interval and level of detail of screenshots to find what might be optimal for a real-time system.

\subsection{Limitations}
Our study has a few limitations. First, due to the complexity and novelty of our study procedure, which involves both passive screenshot collection and three interviews per participant, we slowly rolled out the study and included only 17 participants in the U.S. While our participants are relatively diverse in terms of age and racial identity, a large proportion of them are women. Future work can apply similar methods to a bigger and more balanced sample and in different contexts to examine whether the findings of this study still hold robustly across different population. Second, in the screenshot analysis, we only examined five social media apps. While social media apps are most commonly associated with problematic phone use, and the apps we chose are among the most popular in the U.S., we believe our screenshot analysis technique can be applied to a broader range of apps. For example, in Figure~\ref{fig:proportion-none}, Amazon shopping also showed a high proportion of non-intentional sessions, and online shopping apps also typically have a wide range of passive (e.g., feed for recommended products) and active features (e.g., search). Therefore, future work can investigate fine-grained phone-use behaviors beyond the scope of social media apps and investigate their impact on user's experience. \revise{Finally, since participants could delete entire sessions of screenshots due to privacy concerns, some of their behavior may not be reflected in our data.}

\section{Conclusion}
In this study, we combined experience sampling, surveys, retrospective interviews, and passively collected screenshots analyzed via a multimodal large language model to examine how regret varies depending on user's intention and activity on social media. We found that regret of phone use varies by user intention, and our participants felt most regretful when using their phone without a specific goal. We also found that on social media, participants regretted more when they viewed content that was algorithmically generated and comments or discussion threads and less when they actively communicated with people and searched for information. Additionally, we found that over 60\% of the time people deviated from their intention to communicate to social media browsing, which slightly increased their regret. We argue that designers and policy-makers who seek to improve user's experience and autonomy can assess and measure user's intentionality and reduce features and designs that might funnel users into regrettable use. Our screenshot analysis approach can also be adopted in just-in-time and fine-grained intervention systems that seek to reduce smartphone overuse.

\begin{acks}
Special thanks to all the participants for completing the intensive data-collection process and contributing their insights to this study. We also thank Anind Dey, Je-Wei Hsu, Shahan Ali Memon, Adiba Orzikulova, and Nic Weber for their insightful suggestions at various stages of this paper. Additionally, we thank the anonymous reviewers for their detailed and thoughtful feedback, which helped us refine this work. Alexis Hiniker is a special government employee for the Federal Trade Commission. The content expressed in this manuscript does not reflect the views of the Commission or any of the Commissioners.
\end{acks}

\bibliographystyle{ACM-Reference-Format}
\bibliography{sample-base}

\appendix
\section{LLM Prompts}\label{prompt}
\subsection{Screenshot to Text Description (Step 1 in Figure~\ref{fig:llm})}\label{prompt1}
\texttt{\{"role": "system", "content": "You are a helpful assis\-tant specializing in visual content analysis of smart\-phone screenshots."\}, \\ \{"role": "user", "content": ["Here are two smartphone screenshots, taken 5 seconds apart within the same app Instagram. Your primary task is to describe all of the visual elements (including all of the UI components and content) and the user's activity in the SECOND screenshot in great detail. Please also use the first screenshot as contextual information, and describe how the user transitioned from the first screenshot to the second. Do not describe the first screenshot in detail. In your response, refer to the first screenshot as 'the previous screenshot', and the second screenshot as 'the screenshot'.", \{"image": ...\}, \{"image": ...\}]\}}

\subsection{Text Description to Category (Step 3 in Figure~\ref{fig:llm})}\label{prompt2}
\texttt{\{"role": "system", "content": "You are a helpful assis\-tant specializing in understanding text descriptions of smartphone screenshots."\}, \\ \{"role": "user", "content": "Below are text descriptions of screenshots taken every 5 seconds on Instagram, which include summaries of the visual elements of the screenshot and the user's activity. Your task is to categorize user's activity of the LAST screenshot into one of the following categories. Justify your choices with a step-by-step rationale. \\ \\ Active Communication: The screenshot includes the pre\-sence of a private messaging interface, suggesting the user is actively communicating with specific individuals or groups. \\ Active Search: The screenshot suggests the user is using the search feature to find specific information, content, articles, or items or is consuming content they found through active searching. \\ Consuming Recommendation-Based Content: The screenshot shows explicit indicators of recommendation-based con\-tent in the user's feed, such as a "For You" tab, "Suggested Post" labels, or buttons like "Follow" and "Join" which allow users to subscribe to new content. \\ Consuming Subscription-Based Content: The screenshot shows content already followed or subscribed to by the user in their feed. The screenshot should include indicators such as an active 'Following' or 'Subscrip\-tion' tab of the app, or signs suggesting that the user has already followed the content poster, such as the absence of buttons next to the content poster or community to follow, join, or subscribe in a feed interface. \\ Consuming Content Shared by Others: The screenshot suggests the user is viewing content shared or reposted by someone they followed or opened from a private conversation. \\ Viewing Comments or Discussion Thread: The screenshot suggests the user is viewing the comment section of discussion thread of a social media post. \\ None of the Above: When the user opens a link to a website, when the user sees sponsored content or ad, when it is unclear if user is viewing content posted by someone they followed or recommended to them, or when seeing these screens: home screen, notification screen, a black, dimmed, or blank screen, a screen showing a survey prompt. \\ \\ RULES: \\ 1. Pay attention to how the user transitioned from earlier screenshots. Look for evidence from the four previous screenshots when needed. For example, if the last screenshot shows some content in full screen, without an indicator of where the content came from, look at previous screenshots where the same content appeared and find the relevant indicators. \\ 2. Do NOT always assume that the user is Consuming Subscription-Based Content when you do not see a follow button. Check if the user previously opened the video from a conversation, or did active searching to find the result, and look at previous screenshots to find out if the follow button was present when the user was viewing the same piece of content. \\ 3. Do NOT consider general social media features such as likes, comments, shares, or hashtags as indicators of recommendation-based content. \\ 4. If the user is looking at a post with a comment section but the majority of the screenshot is not about the comments, do not categorize the user's activity as Viewing Comments or Discussion Thread. \\ 5. If the user is looking at multiple social media posts in their feed, pay attention to the one the user is most likely looking at, such as the one in the center and showing the full content. Ignore partially visible posts in the feed. \\ \\ Screenshot Descriptions: \\ Screenshot 1 Description: \{...\} \\ Screenshot 2 Description: \{...\} \\ Screenshot 3 Description: \{...\} \\ Screenshot 4 Description: \{...\} \\ Screenshot 5 Description: \{...\} \\ "\}}

\section{\revise{Interview Codebook}}\label{codebook}
See Tables \ref{tab:codebook}.

\renewcommand{\arraystretch}{1.4}
\begin{table*}
  \caption{\revise{Interview codebook.}}
  \label{tab:codebook}
  \Description{
  This table presents a qualitative coding scheme used for analyzing interview data, categorizing different aspects of phone usage. It consists of four columns: Code, Secondary Code, Definition, and Example. The code “Session Duration” includes secondary codes: Extensive Use, Brief and Intermittent Use. The code “Intended Use” includes secondary codes: No Specific Goal, Having a Specific Goal, Intention-Behavior Inconsistency. The code “Social Media Activity” includes secondary codes: Active Communication, Active Search, Consuming Recommendation-Based Content, Consuming Subscription-Based Content, Consuming Content Shared by Others, Viewing Comments or Discussion Thread. The code “External Factors” includes secondary codes: Bedtime Phone Use, Using the Phone at Work, Using the Phone During Commuting.
  }
  \small
  \begin{tabular}{m{1cm} m{3cm} m{4cm} m{8.5cm}}
    \toprule
    Code & Secondary Code & Definition & Example \\
    \midrule
    \multirow{2}{1.2cm}{Session Duration} & Extensive Use & The participant spent a prolonged period on their phone. &  ``I think it was also about the length—it was just so long. It felt unnecessarily prolonged. So, yes, I regretted that session.'' (P12, F1) \\
    \cline{2-4}
    
     & Brief and Intermittent Use & The participant spent a short time in a given session. & ``I started watching that and scrolling a bit through my homepage. It didn’t get completely out of hand—I wasn’t scrolling for 30 minutes or anything. I kept it relatively controlled.'' (P9, F1) \\
      \midrule

      \multirow{3}{1.2cm}{Intended Use} & No Specific Goal & The participant engaged with their phone habitually or without a clear goal, or engaged in ``doomscrolling''. &  ``I think sometimes it’s almost unconscious to open a messaging app. For example, in the theater, I guess I was maybe expecting a message, so I opened it. But then I just closed it without doing anything because there wasn’t any message.'' (P8, F2) \\
      \cline{2-4}
      & Having a Specific Goal & The participant engaged with their phone with a clear intention. & ``Someone asked me a question about which division my school operates in, so that’s what I was doing. No regret there—it was a very specific intention.'' (P12, F2) \\
      \cline{2-4}
     & Intention-Behavior Inconsistency & Phone use deviates from the user's original purpose. &  ``You could see it as me trying to complete my task, which was finding a video to help me fall asleep, like rain sounds. But I couldn’t get to that until I ended up checking YouTube Shorts beforehand. That part didn’t have any meaning to me'' (P9, F1) \\
     \midrule

    \multirow{6}{1.2cm}{Social Media Activity} & Active Communication & The participant used direct messaging features to communicate or interact with others. &  ``This is WhatsApp with my family and I definitely don't regret it because I wanted to get updates on what my brother and his girlfriend are doing because they're traveling right now and it's, uh, important for me to communicate with my family.'' (P4, F2) \\
    \cline{2-4}
     & Active Search & The participant searched for a specific piece of information. &  ``I wanted to see what deals were available, and I also wanted to find out what restaurants were there, so I could actually get information.'' (P1, F1) \\
     \cline{2-4}
     & Consuming Recommendation-Based Content & The participant viewed social media content algorithmically recommended by the platform. &  ``So random content, like what Twitter gives me or Facebook. Like almost all of it was random. It wasn’t people I knew, right. So like that content, I end up regretting more than if it’s content of like people I know or a work thing, like an article I was reading.'' (P11, F1)  \\
     \cline{2-4}
     & Consuming Subscription-Based Content & The participant viewed social media content posted by accounts they followed. &  ``Every time I see her, I watch her latest video. So that's interesting. It must have been about just feeling like I was spending too much time on social media because I like watching her videos.'' (P7, F1) \\
     \cline{2-4}
     & Consuming Content Shared by Others & The participant viewed social media content shared by their friends. &  ``I opened a TikTok because my friend sent it to me. I don’t regret it because I needed to watch it to reply to her message. It was pretty short, and I did find it entertaining.'' (P4, F1) \\
     \cline{2-4}
     & Viewing Comments or Discussion Thread & The participant viewed comments or discussions from others on various topics. &  ``Sometimes when I go into comment sections, then I'd be more likely to regret the content because especially with political topics, there's so many obnoxious comments in the comment section.'' (P7, F1) \\
     \midrule

    \multirow{3}{1.2cm}{External Factors} & Bedtime Phone Use
 & Phone use during or before bedtime that affected sleep. &  ``When I'm playing on social media at night, when I should be sleeping, when I'm having problems sleeping, tends to be when I regret it more. I realized during this that I feel more regretful about being on social media in the middle of the night, especially when it's just looking at videos and memes.'' (P6, F1) \\
    \cline{2-4}
     & Using the Phone at Work & Phone use during work time. & ``If I'm stressed with work, I think that leads to more regret because like my time which I should be working and then I feel like I wasted my time.'' (P8, F2) \\
     \cline{2-4}
    & Using the Phone during Commuting & Phone use during the time the participant was traveling between places. & ``If I'm on social media while I'm on the bus or light rail or walking to work, I never regret that time because it's not like there's many other productive things I could realistically be doing while I'm commuting.'' (P4, F1) \\
    \bottomrule
  \end{tabular}
\end{table*}

\end{document}